\documentclass[manuscript]{aastex631}

\usepackage{graphicx}
\usepackage{perpage}
\usepackage{amsmath}
\usepackage{color}

\MakePerPage{footnote}

\newcommand{\beq}	{\begin{equation}}
\newcommand{\eeq}	{\end{equation}}
\newcommand{\beqa}	{\begin{eqnarray}}
\newcommand{\eeqa}	{\end{eqnarray}}
\newcommand{\e}	        {$^{-1}$}
\newcommand{\ee}	{$^{-2}$}
\newcommand{\eee}	{$^{-3}$}

\newcommand{\calm}	{{\cal M}}

\newcommand{\avg}[1]    {{\langle #1 \rangle}} 

\newcommand{\vecB}	{{\bf B}}

\newcommand{\alfven}    {{Alfv$\acute{\rm e}$n }}

\newcommand{\muphi}	{\mu_{\Phi}}

\newcommand{\nh}	{n_{\rm H}}
\newcommand{\nhc}	{n_{\rm H,c}}

\newcommand{\vff}       {v_{\rm ff}}

\newcommand{\va}	{v_{\rm A}}
\newcommand{\vrms}	{v_{\rm rms}}

\newcommand{\ml}    {M_{\ell}}

\newcommand{\avir}      {\alpha_{\rm vir}}
\newcommand{\avirf}      {\alpha_{\rm vir,f}}

\newcommand{\lj}	{\lambda_{\rm J}}

\newcommand{\ma}	{{\calm_{\rm A}}}

\newcommand{\mao}	{{\calm_{\rm A,0}}}

\newcommand{\mmug}	{\mu{\rm G}}
\newcommand{\nbh}	{\bar n_{\rm H}}

\newcommand{\nbht}	{\bar n_{\rm H,\, 3}}

\newcommand{\rms}       {{\rm rms}}

\newcommand{\snt}       {\sigma_{\rm nt}}

\newcommand{\Nh}	{N_{\rm H}}
\newcommand{\Nhz}	{N_{{\rm H},z}}
\newcommand{\Nhc}	{N_{{\rm H}c}}
\newcommand{\Nhy}	{N_{{\rm H},y}}

\newcommand{\NH}	{N_{\rm H}}
\newcommand{\NHc}	{N_{{\rm H}c}}
\newcommand{\tff}	{t_{\rm ff}}

\newcommand\cs		{c_{\rm s}}

\newcommand\pc		    {{\rm pc}}

\def\tff				{t_{\rm ff}}

\newcommand{\cpcs}       {{C_{\rm pc}^*}}
\newcommand{\crit}		{{\rm crit}}
\newcommand{\degree}	    {{$^\circ$}}

\newcommand{\get}       {\textit{getsf}}
\newcommand{\wch}       {w_{\rm ch}}

\newcommand{\mlc}       {M_{\ell,\rm crit}}
\newcommand{\mlv}       {M_{\ell,\rm vir}}

\newcommand{\msun}      {M$_\odot$}

\newcommand{\nthr}      {n_{\rm H,thr}}
\newcommand{\rot}       {{\rm rot}}

\newcommand{\vir}		{{\rm vir}}
\newcommand{\wpl}        {W_{\rm P}}

\begin{document}

\title{Magnetized interstellar molecular clouds - III. Filament Collisions and Core Formation: Insights into Substructures and Evolution}

\correspondingauthor{Pak Shing Li}
\email{pakshingLi@shao.ac.cn}

\author[0000-0001-8077-7095]{Pak Shing Li}
\affiliation{Shanghai Astronomical Observatory,
Chinese Academy of Sciences \\
80 Nandan Road, Shanghai 200030, PR China}

\author{Christopher F. McKee}
\affiliation{Depeartments of Physics and of Astronomy, University of California at Berkeley \\
Berkeley, California, USA}

\author{Richard I. Klein}
\affiliation{Department of Astronomy, University of California at Berkeley \\
Berkeley, California, USA}

\begin{abstract}

We present the results of a driven, ideal magnetohydrodynamics simulation of a molecular cloud to study the formation of filamentary structures in molecular clouds and their fragmentation into dense cloud cores.  Shock compression results in the formation of sub-parsec long filamentary structures.  Large-scale supersonic flows push these small filaments together and form massive filamentary clouds.  The main filament that forms stretches across the full 4.4 pc length of the simulation and has a median FWHM of 0.075 pc, similar to that observed in many molecular clouds. Subfilaments inside the main filament have a median FWHM of 0.03-0.04 pc, similar to the ``fibers" observed in molecular lines.   The collision and merger of subcritical subfilaments is an important mechanism for the formation of prestellar cores.  Colliding subfilaments and the resulting core at their intersection is a hub-filament system similar to that associated with massive star formation, but on a smaller scale.  Core mergers can be significant in the growth of pre-stellar cores and may play a role in the formation of binary stars.

\end{abstract}


\section{Introduction} 
\label{sec:intro}

Filamentary structures are ubiquitous in the interstellar medium, ranging from scales much less than 1 pc to kpc scales.  Long, massive filamentary structures are commonly observed inside giant molecular clouds (GMCs) in molecular emission lines and in dust emission and absorption \citep{hacar23,pineda23}.  High-column filamentary molecular clouds, such as SDC13 \citep{peretto2014} and IC5146 \citep{arzoumanian11}, qualify as Infrared Dark Clouds, which are generally defined to have hydrogen column densities $\Nh>2\times 10^{22}$~cm\ee\ \citep[e.g.,][]{simon2006,henshaw2013}. Peak column densities in the filaments in nearby molecular clouds are in the range just below this, $6\times 10^{21}\mbox{ cm\ee}\la\Nh\la 2\times 10^{22}$~cm\ee\ \citep{arzoumanian19}.  
  
Observations on different size scales show that filamentary molecular clouds have a hierarchical structure \citep[see the reviews in][]{hacar23,pineda23}.  Very large filamentary clouds, such as Orion A \citep{yun2021}, are found to be composed of many shorter, denser filamentary clouds of a few to more than 10 pc in length.  In the northern part of Orion A, the Integral-Shaped Filament is composed of many shorter filaments about a parsec long, which in turn are composed of even thinner and shorter substructures \citep{suri2019, zhang2020}.  
This hierarchical structure is seen in most nearby molecular clouds. \citet{pineda23} summarize the evidence that there is a typical FWHM of filaments in nearby molecular clouds of $\simeq 0.1$~pc, 
Inside the filaments in some, but not all, molecular clouds, narrower velocity-coherent structures (``fibers") have been observed, with widths in the range $0.02-0.1$ pc \citep{hacar23}.  The mass per unit length of these fibers is comparable to the value needed for thermal pressure to support them against radial collapse due to self-gravity.
There is often more than one fiber in every cross section along an entire cloud--in effect, the cloud is a bundle of subparsec-long filamentary substructures.  In some locations, fibers are observed to be moving supersonically relative to each other, with relative velocities of 1-2 km s$^{-1}$ \citep[e.g.][]{hacar13}.

How do these filamentary substructures form inside molecular clouds?  
As discussed in \citet{hacar23} and \citet{pineda23}, the hierarchical structure could be the result of a top-down fragmentation process and/or a bottom-up process in which turbulence creates small filaments that are then swept together and accrete more material. How do all the filamentary structures interact in a very turbulent environment?  Dense cores are observed along some substructures \citep[e.g.][]{tafalla2015,konyves2015,Yang2024}; how exactly do cores form in the substructures?
\citet{hacar2017} suggested that collision of fibers in a crowded environment will be unavoidable and would lead to core formation.

In \citet{li19} (hereafter Paper II), we began to address these questions by simulating the structure of a 4.55 pc cube in a typical Giant Molecular Cloud (GMC). We found that a significant fraction of the mass was concentrated in a filament that stretched across the computational domain of about 4.4 pc in length; we refer to this as the ``main filament".  The width of the main filament, defined in Paper II as having a column density $\Nh>2\times 10^{22}$~cm\ee, is about 0.25 pc.  Inside the main filament, there is a large number of subparsec-long filamentary substructures, which we term ``subfilaments".  Filaments and subfilaments are required to have aspect ratios $>3$. Numerical simulation allows us to visualize the formation of these filamentary substructures, how they interact 
inside the filamentary cloud, and how dense cores form as a result of their interaction. When observed in projection, different subfilaments can appear to be part of a single filamentary component of the main filament. This ambiguity can often be resolved with the addition of velocity information; filamentary substructures in PPV space are termed ``fibers" \citep{hacar2017}. There are two special cases: Two fibers correspond to a single subfilament when two subfilaments collide and form a single structure with a sharp internal velocity gradient. Two subfilaments correspond to a single fiber when two approximately parallel subfilaments have the same velocity. With these two exceptions, fibers and subfilaments are identical structures, provided the dust-to-gas ratio and the CO emissivity are about constant. Note that even if these structures are physically identical, their measured properties could be different if different methods are used to analyze them.  In Section \ref{sec:filaments} we discuss how the main filament is similar to the filaments studied in a series of papers beginning with \citet{arzoumanian11} and the subfilaments are related to the ``fibers" studied in a series of observational papers beginning with \citet{hacar13} and in the simulation paper of \citet{zamora2017}.

The paper is structured as follows: The numerical method is explained in Section \ref{sec:sim}.  The stability of gaseous filaments is reviewed in Section \ref{sec:filament_stab}. In Section \ref{sec:filaments}, we discuss the formation and evolution of subfilaments.  Section \ref{sec:filaments} has four subsections, where we describe how to determine the FWHM of the main filament in the simulation, the statistics of physical properties of subfilaments using 2D column density maps, the evolution of a subcritical subfilament to become supercritical in a crowded environment using 3D data, and the effects of helical magnetic field to a subfilament.  
In Section \ref{sec:core}, we discuss core formation by subfilament collisions, the growth of dense cores by mergers, and the core mass function.  In section \ref{sec:two}, we report in detail on the evolution of physical properties of two cores.  We use the first core as an example to discuss both the effects of shocks and the energy balance around the core and the attached subfilaments.
The conclusions of this work are given in Section \ref{sec:conclusion}. Four animations from the simulation are included in this paper to enable visualization of the interaction among subfilaments and cores.

\section{SIMULATION PARAMETERS AND METHODS} \label{sec:sim}

We use our multiphysics adaptive mesh refinement (AMR) code \textsc{Orion2} \citep{li12a,li21} to perform a large-scale simulation to study the formation and structure of filamentary clouds. The full version of \textsc{Orion2} is capable of solving ideal MHD along with coupled self-gravity \citep{martin08}, radiation transport, and feedback physics. For the simulation reported in this paper, we study the formation of subfilaments and cores in the absence of star formation. Thus, radiation transport and feedback physics are ignored. Our simulation is an ideal MHD, driven turbulence simulation with gravity turned on after the turbulence has settled into a steady state.

Simulations of isothermal MHD turbulence are scale-free and are defined by two dimensionless parameters: the 3D thermal Mach number, $\calm = \vrms/c_s$, and the \alfven Mach number, $\ma = \vrms/\va$, where $\vrms=\surd 3 \snt$ is the mass-weighted rms nonthermal velocity dispersion, $\snt$ is the mass-weighted 1D non-thermal velocity dispersion, $c_s = 0.188$ km s$^{-1}$ is the isothermal sound speed at 10 K, and $\va$ is the mass-weighted \alfven velocity.  Self-gravity introduces a third dimensionless parameter, the virial parameter \citep{bertoldi1992},
\beq
\avir\equiv \frac{5\sigma^2 R}{GM}
\label{eq:avir}
\eeq
where the size of the simulation box is related to $R$ by $\ell=2R$ and $\sigma =(\snt^2+c_s^2)^{1/2}$ is the total 1D mass-weighted velocity dispersion.  

To set the scales in the simulation, we need three dimensional parameters.  One is the isothermal sound speed, $c_s$, which sets the velocity scale, and a second is the gravitational constant, $G$, which sets the scale for $M/\ell$.  For the third dimensional parameter, we adopt the coefficient, $C$, in the turbulent line-width-size relation observed in supersonic flows in GMCs and seen in simulations of supersonic turbulence, $\snt=C R^{1/2}$,
where $\snt$ is averaged over a sphere of radius $R$ \citep{mckee07}.
When $R$ is measured in pc, this becomes $\snt=C_\pc R_\pc^{1/2}$; in the Galaxy, $C_\pc=0.72$~km~s\e. Normalizing to the Galactic value, we have
\beq
\snt=0.72\,\cpcs R_{\rm pc}^{1/2}~~~\mbox{km s\e}=0.72\,\cpcs(\ell_{\,\pc}/2)^{1/2}~~~\mbox{km s\e},
\label{eq:lws}
\eeq
where $\cpcs = C_\pc/(0.72$~km~s\e\ pc$^{1/2}$) allows deviations from the typical linewidth-size relation.  
Since $\snt=(\calm/\surd 3)\cs$, where $\calm$ is the 3D Mach number, and under the assumptions that the gas is fully molecular and highly supersonic on large scales, this allows us to express the size of the simulation box in terms of $\cs$ and $C$ as
\beq
\ell_\pc= \frac 23\;\frac{\calm^2 \cs^2}{C_\pc^2}=0.0455\left(\frac{\calm^2 T_1}{\cpcs^2}\right),
\label{eq:lscale}
\eeq
where $T_1=T/(10$~K). The 3D sonic length is then this expression evaluated at $\calm=1$: $\ell_s=0.0455T_1/\cpcs^2$ pc.

Under the same assumptions of highly supersonic flows and fully molecular gas, the other physical parameters of the turbulent system--
the flow time (or crossing time), $t_f$, the mean density of H nuclei, $\nbh$, the mass of the box, $M$, and the column density,
$\NH=\nbh\ell$---can
then be expressed as (see the Appendix in \citealp{mckee10}).:
\begin{eqnarray}
t_f&=&\frac{\ell}{\vrms}=2.36\times 10^5\left(\frac{\calm T_1^{1/2}}{\cpcs^2}\right)~~~\mbox{yr},
\label{eq:tscale}\\
\nbh&=&9.6\times 10^4\left(\frac{\cpcs^4}{\avir\calm^2T_1}\right)~~~\mbox{cm\eee},
\label{eq:nscale}\\
M&=& 0.311\left(\frac{\calm^4T_1^2}{\avir\cpcs^2}\right)~~M_\odot,
\label{eq:mscale}\\
\NH&=&1.34\times 10^{22}\left(\frac{\cpcs^2}{\avir}\right)~~~\mbox{cm\ee}.
\label{eq:Nscale}
\end{eqnarray} 
The initial value of the uniform magnetic field along the $z$-axis is given by
\beq
B_0=4.56\;\left(\frac{\nbht T_1}{\beta_0}\right)^{1/2}\mmug
=31.6\left(\frac{\cpcs^2}{\avir^{1/2}\ma}\right)\mmug.
\label{eq:bscale}
\eeq
where $\nbht=\nbh/(1000$~cm\eee), and $\beta_0=8\pi\rho \cs^2/B_0^2$ is the initial plasma beta.

\begin{table}
\caption{Initial Physical Properties of the Entire Region}
\label{tab:region}
\begin{tabular}{lc}
\vspace{-0.3cm}\\
\hline
\hline
3D thermal Mach number ($\calm$)  & 10           \\
Virial parameter ($\alpha_{\rm vir}$) & 1        \\
\alfven Mach number ($\mao$)   & 1               \\
Plasma beta ($\beta_0$)        & 0.02            \\
Mass-to-flux ratio ($\mu_{\Phi}$)     & 1.62     \\
Temperature ($T$)              & 10 K            \\
Total mass ($M$)               & 3110 $M_{\odot}$\\
Size ($\ell$) \tablenotemark{a} & 4.55 pc         \\
Mean density ($\nbh$)    & 960 cm$^{-3}$   \\
Mean column density ($\bar{N}_{\rm H}$) & $1.34\times 10^{22}$ cm\ee\\
1D velocity dispersion ($\sigma$) & 1.1 km s\e   \\
Mean magnetic field ($B_0\mathbf{\hat z}$)    & 31.6 $\mu$G     \\
Flow time ($t_f$)             & $2.36\times10^6$ yr \\
Free-fall time ($t_{\rm ff}$)  & $1.40\times10^6$ yr \\
\hline
\end{tabular}
\tablenotetext{a}{Maximum resolution is 2048 grid cells, or $2.2\times 10^{-3}$ pc in the computational box of size $\ell=4.55$ pc.  The run time is 894 kyr $=0.64\tff$}.
\end{table}
The simulation has $\calm=10$, $T=10$~K, $\avir=1$ and $\cpcs=1$, so that the size of the simulated turbulent region is $\ell = 4.55$ pc, the mean number density $\nbh = 960$ cm$^{-3}$, the total mass $M = 3110 \,M_{\sun}$, and the mean column density of the system $\NH = 1.34\times 10^{22}$ cm$^{-2}$. The free-fall time of the entire turbulent region based on the mean density,
\beq
\tff=\left(\frac{3\pi}{32G\bar\rho}\right)^{1/2}=1.37\times 10^6\nbht^{-1/2}~~~\mbox{yr},
\label{eq:tff}
\eeq
is $\sim 1.4\times10^6$ yr, which is $0.59 t_{\rm f}$. The magnetic field strength in the simulation is moderately strong, 
with $\mao=1$, corresponding to an initial plasma $\beta_0 = 0.02$. The initial simulation parameters are summarized in Table \ref{tab:region}.

The relative importance of gravity and magnetic fields can be measured by the normalized mass-to-flux ratio, $\mu_\Phi=M/M_\Phi$, where the magnetic critical mass is given in terms of the magnetic flux threading the cloud, $\Phi$, by $M_\Phi=\Phi/(2\pi G^{1/2})$.
Insofar as the magnetic field is frozen in the matter, 
gravity can overcome the field for $\mu_\Phi>1$, whereas gravitational collapse is impossible for $\mu_\Phi<1$.
For a cubical box, $\mu_\Phi$ is related to the other dimensionless parameters by $\mu_\Phi=(5\pi/6\avir)^{1/2}\mao=1.62$ \citep{mckee10}, which is only slightly supercritical; this corresponds to a moderately strong magnetic field.

In the first crossing time, we drive turbulence in the system at the largest scales with wave numbers $k = 1 - 2$ on a single base grid of $512^3$ using the procedure described in \citet{maclow99}.  In the second crossing time, we allow two levels of refinement with a refinement ratio of 2 at each fine level.  In addition, the volume coverage of the fine levels is maintained to be about 12\% of the volume coverage of the coarser level \citep{li19}.
Since the size of the entire region is $\ell = 4.55$ pc, the finest resolution corresponds to a cell size $\Delta x=2.22\times 10^{-3}$~pc = 460 au, sufficient for the study of the subfilaments of width of order 0.02 pc observed by \citet{hacar18}.

At the end of the second crossing time, we turn on gravity and continuously drive the system with a constant energy injection rate that maintains the entire system at Mach 10.  After gravity is activated, we include the additional refinement criterion that the cell size be no more than a fraction $J$ of the Jeans length, $\Delta x<J\lj$ \citep{truelove97}. We set $J=1/8$ so that the Jeans length is resolved by at least 8 cells.  This sets a limit on the density, $\rho<\pi J^2\cs^2/(G\Delta x^2)$. For $T=10$~K and $\Delta x=2.22\times 10^{-3}$~pc for the finest level of refinement, this corresponds to a resolution limit on the density of hydrogen nuclei of $\nh<2.4\times 10^6$~cm\eee.
In our studies of star formation \citep[e.g.][]{li18}, we introduced a sink particle at this point to represent a star that has formed. We do not do that here since we are not following star formation; instead, we allow gas to accumulate in a small number of cells, which therefore acquire unresolved densities and pressures.
We adopt periodic boundary conditions and assume an isothermal equation of state.  We set the time $t=0$ at the moment when we turn on gravity and then run until $t=0.64\tff=894$ kyr.

The simulation was run with self-gravity for about $0.64\tff$ ($\sim 900$ kyr). 
The total computational cost of this simulation on 4096 processors is 1.8 million CPU hours, including the initial two crossing times driving without gravity.
\begin{figure*}
\begin{interactive}{animation}{A1.mp4}
\includegraphics[scale=0.75]{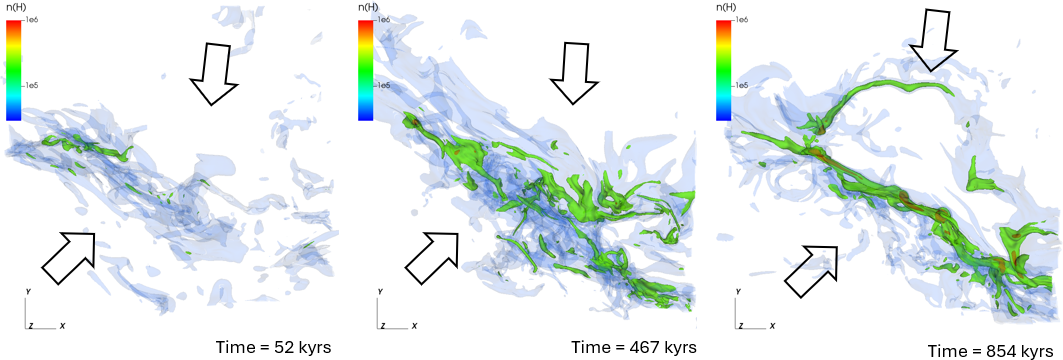}
\end{interactive}
\caption{Three snapshots (1 pc length box) of an enlarged portion  of the main filament simulated at three different times, showing the internal subfilaments at different densities (blue to yellow) and dense cores (red).  The white arrows indicate the direction of the large-scale turbulent flow.  See animation A1 for visualizing the evolution of this part of the main filament starting from the time when gravity is activated.  \it{The animation associated with this figure is available in the online version of the published article on ApJ.}
\label{fig:filament_cloud}}
\end{figure*}

In Figure \ref{fig:filament_cloud}, we show three snapshots (1 pc length box) of the evolution of an enlarged portion  of the main filament in the simulation after gravity is activated (see Figure 4 in Paper II). The main filament is located in a converging flow in the large-scale turbulence.  Before gravity is activated, many low-density subparsec-long filaments (blue) already exist everywhere in the entire simulation region.  They are pushed together by the large-scale flow to the converging locations.  The low-density subfilaments in the main filament collide and merge, becoming denser (green) and longer.  Later, some very dense subfilaments become unstable and fragment into dense cores (red) at different locations along subfilaments or at their intersections (see Section \ref{subsec:filament_evol} for an example of the evolution of a subfilament).  This process is shown in the animation A1 (Figure \ref{fig:filament_cloud}).  In the next section, we discuss the theoretical stability of gaseous filaments.

\section{Stability of gaseous filaments \label{sec:filament_stab}}

A gas filament in a molecular cloud is subject to supersonic turbulence and gravitational forces in a magnetized environment.  If there are protostars nearby, it will be subject to radiative feedback as well.  Here we briefly review the stability of filaments in the absence of nearby protostars.  Theoretical analysis of the stability of filaments has focused on the case of axisymmetric filaments, and we assume asixymmetry here. For an unmagnetized gas filament, the maximum mass per unit length that is stable against radial gravitational collapse, $M_{\ell,\crit}$, is the virial line mass, $M_{\ell,\vir}$, at which the gravitational self-energy is twice the internal energy due to motions normal to the axis \citep{ostr64,fiege00a},
\beq
M_{\ell,\vir}=\frac{2\sigma^2}{G},
\label{eq:virlmh}
\eeq
where $\sigma$ is the total mass-weighted 1-D velocity dispersion as noted above.  The virial parameter of the filament is
\beq
\avirf=\frac{M_{\ell,\vir}}{\ml}=\frac{2\sigma^2}{G\ml},
\label{eq:avirf}
\eeq
where $\ml$ is the actual line mass of the filament.  
Unmagnetized filaments with $\avirf>1$ ($\ml<\mlv$) are stable against radial collapse; those with $\avirf< 1$ ($\ml>\mlv$) will collapse. Unmagnetized filaments with $\avirf<2$ have negative total energy and are gravitationally bound.  For purely thermal support, the critical line mass for the onset of collapse is $\mlc=16.4(T/10\mbox{ K})$ \msun\ pc\e.  Many observed gas filaments are found to have line masses in excess of this critical value \citep[e.g.][]{palm13,li22,chung2023}.  If the filaments are not collapsing, turbulent motions and/or magnetic fields must provide the additional support.  Note that if the gas has an adiabatic index $\gamma\neq 1$ so that the temperature can change when the density changes with time, then a filament with $\gamma>1$ will heat upon contraction so that $M_{\ell,\vir}$ will increase and the degree of radial collapse will be limited; only filaments with $\gamma<1$ can collapse to very high densities \citep{lars05}.  

In the presence of a magnetic field, the critical line mass depends on the structure of the field.  In the case where there is a uniform magnetic field parallel to a filament of constant density, the results of \citet{fiege00a} imply that the critical line mass is
\beq
M_{\ell,\crit}=\frac{2\sigma^2+v_{{\rm A},z}^2}{G},
\label{eq:virlmb1}
\eeq
where the filament is oriented along the $z$ axis and $v_{{\rm A},z}$ is the \alfven\ velocity for the axial field. 
Whereas axial magnetic fields always increase the critical line mass, thereby working to stabilize filamentary clouds, helical fields squeeze the filament and reduce the critical mass (cf. \citealp{fiege00a}),
\beq
M_{\ell,\crit}=\frac{2\sigma^2+v_{{\rm A},z}^2}{G+v_{{\rm A},\phi}^2/\rho r^2},
\label{eq:virlmb2}
\eeq
where $r$ is the cylindrical radius and it has been assumed that $\rho$ and $B_\phi/r$ are constant.
The toroidal component is now working with gravity to squeeze the filament, thereby reducing the critical line mass.  

Observations show that the orientation of the magnetic field changes from parallel to filamentary gas structures to perpendicular as the column density exceeds $\Nh\sim 10^{21}$~cm\ee\ \citep{plan16}. The column density of the dense filament in our simulation exceeds this threshold, and indeed we find that the field usually is roughly perpendicular to the filament.  In the perpendicular case, \citet{kashiwagi21} found that the critical line mass for $\gamma\rightarrow 1$ is
\beq
M_{\ell,\crit}\simeq \left(M_{\ell,\Phi}^2+M_{\ell,\vir}^2\right)^{1/2},
\label{eq:crit}
\eeq
where $M_{\ell,\Phi}=\Phi/(2\pi G^{1/2}L)$ and $\Phi=2r_0 L B_\rms$ is the magnetic flux threading the filament.  Here $2r_0$ is the width of the filament.

\citet{kashiwagi21} had a factor of 0.85 before $M_{\vir,\ell}^2$, but we have omitted this factor in order to make the result exact at the limit $\Phi_\ell=0$ \citep{li22}. Although this result was derived under the assumption of axisymmetry, we anticipate that it will remain approximately valid if the filament is flattened along the field lines since the expression for $\mu_\Phi$ is approximately the same for a cylindrical filament as it is for a thin disk \citep{li22}.  Equation (\ref{eq:crit}) implies \citep{li22}
\beq
\frac{M_\ell}{M_{\ell,\crit}}=\frac{1}{\left(\muphi^{-2}+\avirf^2\right)^{1/2}},
\label{eq:mellcr}
\eeq
where $\mu_\Phi=(2\pi G^{1/2}M)/\Phi$ is the ratio of the mass-to-flux ratio to the critical value--i.e., in the flux-freezing approximation, gravitational collapse is possible only for $\mu_\Phi>1$.

\section{Formation and evolution of subfilaments in the main filament}
\label{sec:filaments}

In Paper II we analyzed the same simulation that we studied here and found that a significant amount of mass was concentrated in a single large filamentary cloud that extended over the length of the simulation box.  We concluded that large-scale converging flows in the turbulence first create and assemble many subparsec-long filaments together before gravity is activated.  After gravity is activated, these structures slowly bind together to form a long filamentary cloud.  The subparsec-long filaments evolve into subfilaments of the main filament.   The goal of Paper II was to simulate the formation of an Infrared Dark Cloud (IRDC), which has a column density $\Nh>2\times 10^{22}$~cm\ee\ \citep[e.g.,][]{peretto2014}.  Therefore, we defined the main filament in Paper II as being composed of gas with $\Nh>2\times 10^{22}$~cm\ee, which is a factor 1.5 times higher than the mean column in the simulation box, $1.34\times 10^{22}$~cm\ee.

In Figure 16 in Paper II, the main filament is oriented approximately along the $x$-axis (within about 10\degree\ in the $y$-projection 
and within $30$\degree\ in the $z$-projection; recall that the initial magnetic field was oriented along the $z$-direction). A secondary massive filament is right next to the main filament, making an angle of almost 90\degree\ in the $z$-projection at half a free-fall time (see Fig. \ref{fig:map+skeletons}).
Although the envelope contours of $\Nh>2\times 10^{22}$~cm\ee\ of the two filaments are closely spaced, the $z$-projection provides sufficient separation between the two filaments for a reasonable estimate of the mass and projection area of the main filament.  In the other two projections, the density contour encloses part or all of the secondary filament.  Therefore, in Paper II, the measurement of the physical properties of the main filament is based on the $z$-projection.  The mass and magnetic flux of this filament were based on lines of sight above the IRDC column-density threshold; the average width of the filament is 0.25 pc.  From the estimated mass and the projected length shown in Table 2 in Paper II at half a free-fall time, the main filament has a line mass of 107 \msun\ pc$^{-1}$, 6.5 times the thermal critical value, $2c_s^2/G$, but close to $2\sigma^2/G = 111$ \msun\ pc\e\ due to the mean 1D velocity dispersion of 0.49 km s$^{-1}$; the resulting virial parameter is $\avirf=1.04$.  The mean magnetic field strength of $52\;\mu$G corresponds to $\mu_\Phi=2.78$, so that the field does not contribute significantly to the support. Overall, the main filament is marginally stable against radial collapse, with $M_\ell/M_{\ell,\crit} = 0.91$.  The main filament is not expected to undergo gravitational collapse since $\avirf$ is slightly above unity \citep{fischera12}, but internally some subfilaments are in the process of doing so (see Section \ref{subsec:filament_evol}).

\subsection{The FWHM of the main filament
\label{subsec:fwhm}}

Following \citet{arzoumanian11}, we fit the column density profiles with a Plummer-like expression, which follows from a Plummer-like density distribution,
\beqa
\nh&=&\frac{\nhc}{\left[1+(r/r_f)^2\right]^{p/2}},\\
\Nh&=&\frac{\NHc}{\left[ 1 + (r_p/r_f)^2\right]^{(p-1)/2}},
\label{eq:plummer}
\eeqa
where $r$ is the cylindrical radius, $r_p$ is the projected radius (i.e., the distance to the peak of the projected column-density profile), and $r_f$ is the characteristic radius at which the profile flattens. Only for $p=4$ is such a filament in hydrostatic equilibrium \citep{stod63, ostr64}; observed filaments generally have $p\simeq 2$ though.\footnote{It should be noted that there is a solution for an almost isothermal filament in hydrostatic equilibrium with $p\simeq 2$: \citet{naka02} have pointed out that a singular polytropic filament has $p=2/(2-\gamma)$, where $\gamma$ is the adiabatic index. If $\gamma\rightarrow 1$ from below, then $p\rightarrow 2$ from below, with the mass/length near the axis remaining finite.}  The central column density is related to the central density by $\NHc=n_{{\rm H}c}r_f {\rm B}[1/2,(p-1)/2]$, where ${\rm B}(x,y)$ is the beta function \citep{palm13}. The FWHM of this distribution is
\beq
\wpl=2r_f\left[2^{2/(p-1)}-1\right]^{1/2},
\label{eq:wpl}
\eeq
\citep{schuller21}, which they denote as $D_{\rm HP}^{\rm Plummer}$. For a typical value of $p=2$, this is $\wpl=2\surd 3 r_f=3.46 r_f$.  The mass per unit length of a Plummer-like filament for $p=2$ is
\beqa
M_\ell(r)
&=&\pi r_f^2\rho_c\ln\left[1+(r/r_f)^2\right],\\
&=&9.05 n_{{\rm H}c,5}\left(\frac{\wpl}{0.1\mbox{ pc}}\right)^2 \ln\left[1+(r/r_f)^2\right]~~~\mbox{\msun\ pc\e},
\eeqa
where $\rho_c$ is the central mass density and $n_{{\rm H}c,5}=\rho_c/(10^5\times 2.34\times 10^{-24}$ g) is the central density of H nuclei.\footnote{Some workers fit Gaussians to filament profiles, $\NH=\NHc\exp(-r^2/2s^2)$, so that the FWHM is $W_{\rm G}=(8\ln 2)^{1/2} s$ and $M_\ell\rightarrow 2\pi n_{{\rm H}c}s^2=39.2n_{{\rm H}c,5}(W_{\rm G}/0.1\mbox{ pc})^2$ \msun~pc\e\ for $r\gg s$.}  

For the main filament, we determine the FWHM using two methods.  For both methods we analyze the filament in the $x$-$y$ and $x$-$z$ planes, since the filament is roughly along the $x$-axis.  The computational cell size at the highest AMR refinement level is $2.2\times10^{-3}$ pc, which is the pixel size used for the maps.  To determine the location of the filament spine in Method 1, we make plots of $\Nh$ along the $y$- and $z$-axis for each pixel in the $x$-direction for the entire filament.  For each plot we choose the pixel containing the maximum column density as being in the spine of the filament.  Connecting these pixels gives the spine of the filament.  Discontinuities in the spine break the main filament into different pieces, or segments, along the filament (Figure 2).  Along each segment, a radial profile of the main filament perpendicular to the spine can be created at each pixel from the column density map.  Note that the spines are not exactly along the $x$-axis, similar to the skeletons plotted in the left panel of Figure 2.  Therefore, the column density profiles, which are perpendicular to the spine, are generally not exactly parallel to the $y$- or $z$-axis.  We aligned the profiles so that the peak of each profile was located at the same point, thereby creating a combined radial profile of the main filament.  Since many subfilaments occur in almost every profile along the main filament, the combined profile of the main filament blends subfilaments together and is correspondingly broader than that of the subfilaments.  We fit the combined radial profile using a Plummer-like profile (Eq. \ref{eq:plummer}). At $0.5\tff$, we obtain $\wpl = 0.071$ pc in the $z$ projection (i.e., from data projected onto the $x$-$y$ plane) and 0.070 pc in the $y$ projection (data projected onto the $x$-$z$ plane), which are slightly less than the mean value observed in nearby molecular clouds, 0.1 pc, but within the interquartile range of 0.06-0.13 pc \citep{arzoumanian19}. The width of the main filament in the $x$-$y$ plane above the IRDC threshold, $\Nh>2\times 10^{22}$~cm\ee, is about 0.25 pc.  As discussed above, it is difficult to measure the width in the $y$ direction due to the presence of another filament.

\begin{figure}
\includegraphics[scale=0.75]{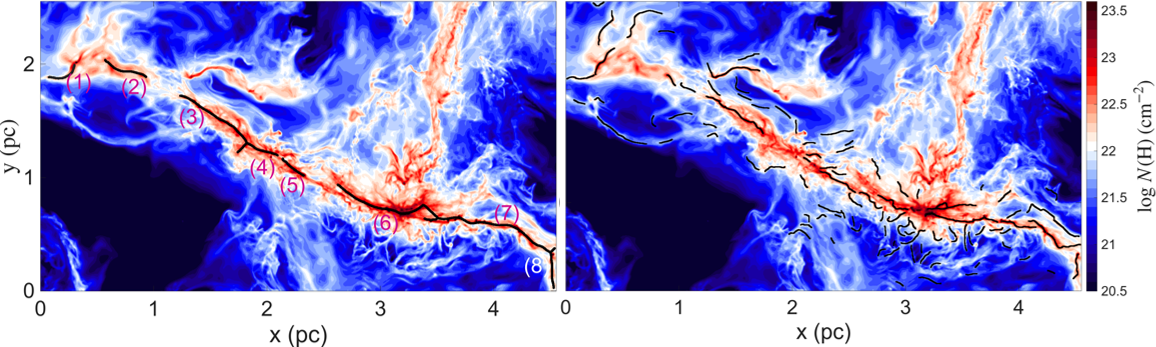}
\caption{The black-colored skeletons identified by \textit{getsf} using $\wch=0.15$ pc (left) and 0.02 pc (right) are plotted on the column density map in the $z$-projection at half the free-fall time, $t=0.5\tff=$ 700 kyr.  The labels of the segments of the main filament in the left panel correspond to the numbering of the column density profiles in Figure \ref{fig:profiles}.
\label{fig:map+skeletons}}
\end{figure}
For the second method, we use a popular software \textit{getsf} \citep{menshchikov2021} for filament identification to obtain the column density profile of the main filament.  The procedure of \textit{getsf} can be summarized as: image preparation, decomposition, flattening of noise and background, detection of filaments, and measurement.  \textit{Getsf} can detect core-like features (termed "sources" in \textit{getsf}) and filaments in the image; here we focus on the detection of filaments.  We smoothed the map using a beam size of 3 pixels so that the beam size is $6.6\times10^{-3}$ pc.  A key input parameter needed by \textit{getsf} in the decomposition is the characteristic width scale of the filament structures to be extracted. \citet{menshchikov2021} terms this the "maximum size of filaments," but we find that filaments can be several times larger than this size scale, so we term it the characteristic width, $\wch$. The background is defined as the smooth density (or intensity) distribution on spatial scales larger than four times $\wch$ that remains after the complete removal of all filaments \citep{menshchikov2021}.  A background is defined for each pixel. For example, for $\wch=0.15$ pc, the background for each pixel is smooth on all scales from 0.6 pc to 4.55 pc.  Filaments are detected using the Hilditch algorithm \citep{hilditch1969}, which determines the skeletons of 2D shapes.  Each pixel of the skeletons has the ratio of density (intensity) to background fluctuations larger than the detection significance required by the users, which is recommended to be 2 and is used in our study.  This results in most skeletons having a density at least twice that of the background.  The physical properties of filaments, such as the FWHM and line mass, are then given by \textit{getsf}.
\begin{figure*}
\includegraphics[scale=0.6]{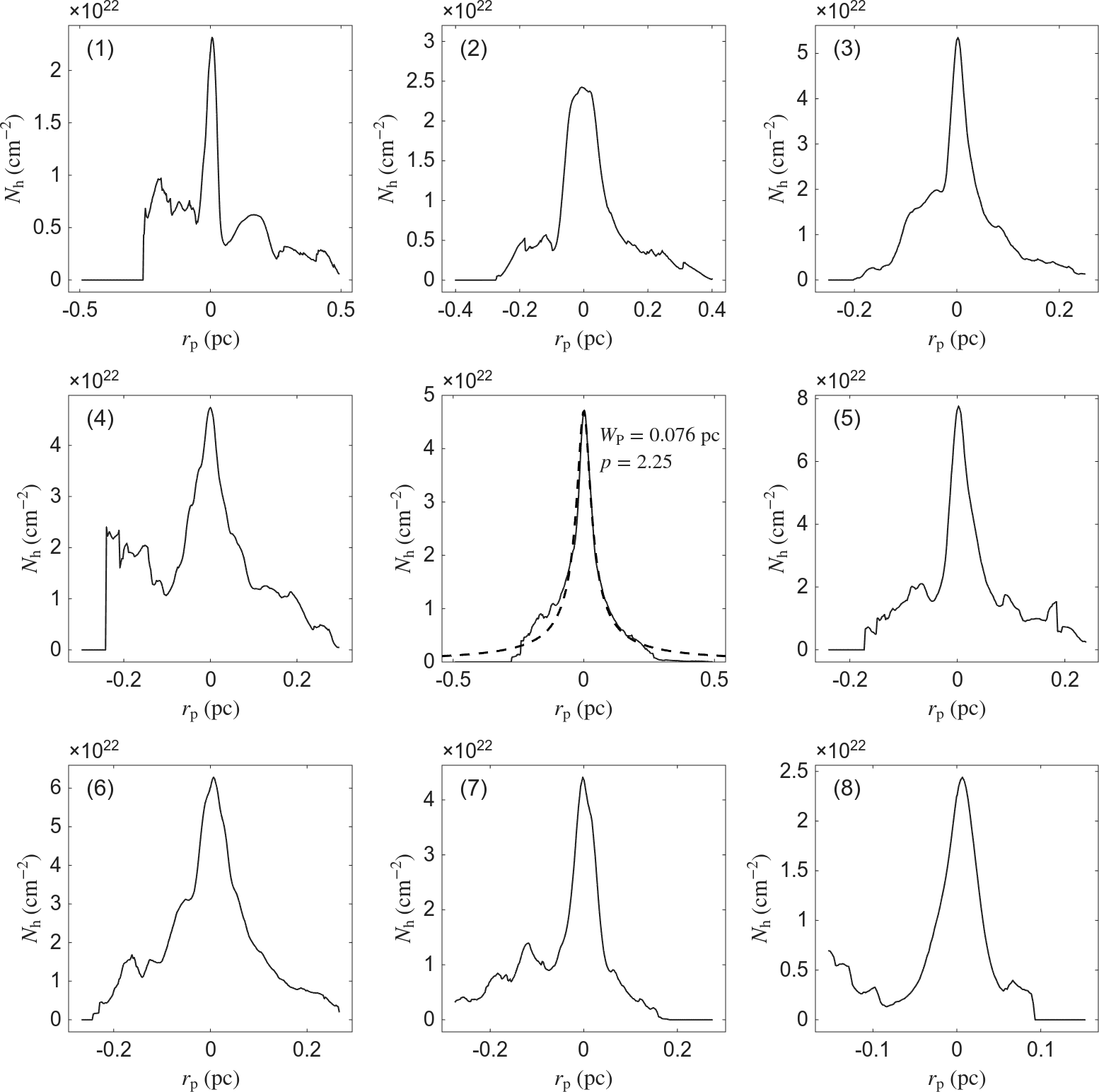}
\caption{The column density profiles normal to the skeleton in $z$ projection at $t=0.5\tff$ averaged along each of the eight segments of the main filament cloud shown in the left panel of Figure \ref{fig:map+skeletons}, which has $\wch=0.15$ pc.  The panel number here is the segment number in Figure \ref{fig:map+skeletons}.  The panel in the middle is the average profile of all eight segments and fitted with a Plummer profile (dashed curve) with the best fitted FWHM $\wpl$ and the index $p$ shown in the figure panel.
\label{fig:profiles}}
\end{figure*}

\begin{figure}
\includegraphics[scale=0.5]{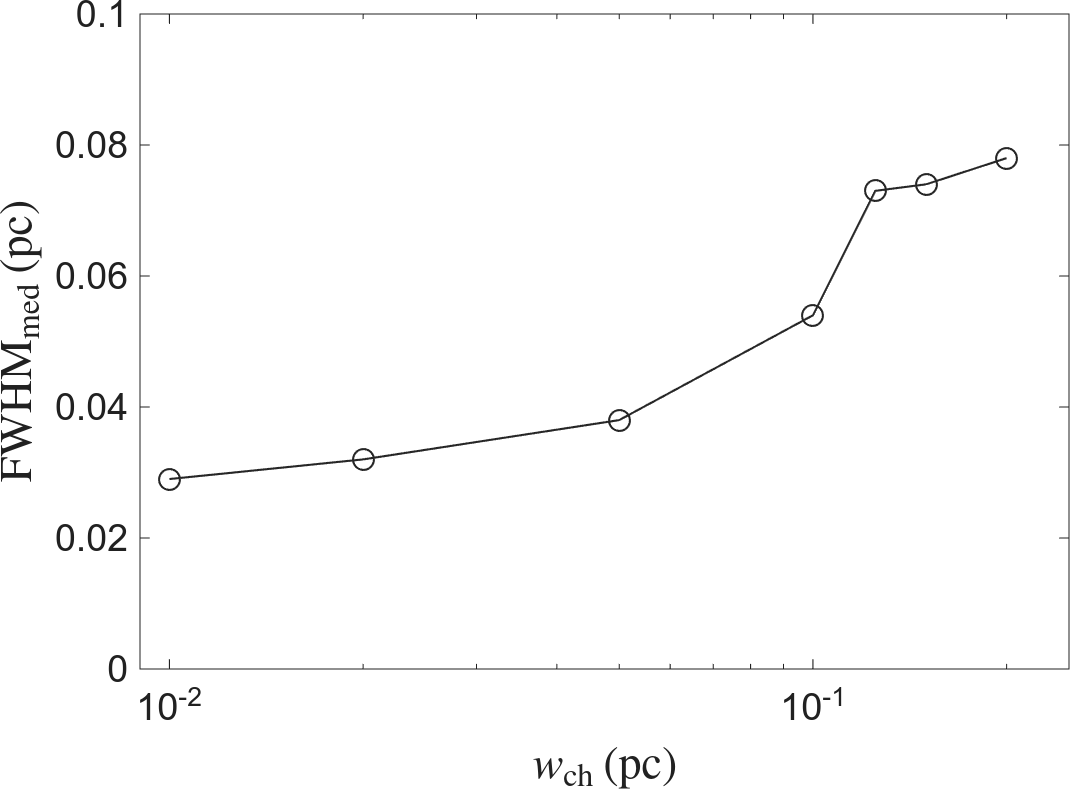}
\caption{The median FWHM of features identified by \textit{getsf} as a function of $\wch$. For small $\wch$, these are subfilaments; for $\wch\ga 2W_{\rm P}=0.15$, they are segments of the main filament.
\label{fig:fwhm_wch}}
\end{figure}

We used $\wch = 0.15$ pc, twice as large as the FWHM of the main filament, to find the skeletons and corresponding segments of the main filament.  The lengths and values of the FWHM of the skeletons were provided by \textit{getsf}.  Similarly, the radial profiles for obtaining FWHM by \textit{getsf} are also perpendicular to the skeletons.  At $t=0.5\tff$, eight segments meet the aspect ratio requirement; these are plotted in the left panel of Figure \ref{fig:map+skeletons}.  
The radial profiles of these eight segments in the $z$-projection are shown in Figure \ref{fig:profiles}.  Observe that some of the segments appear to have substantial subfilaments that are separate from the main feature in the profiles (e.g., segments 1, 4 and 8); however, \get\ identifies features based on skeletons, and treats each segment as a single structure.  We then combined these eight profiles together. Fitting a Plummer function to this profile gives FWHM = 0.076 pc, whereas direct evaluation of the full width at half maximum of the combined profile gives 0.074 pc. 
In the $y$-projection (i.e., as a function of $z$ in the $x$-$z$ plane), the best fit Plummer function to the combined profile gives a FWHM of 0.072 pc, whereas direct evaluation of the FWHM of the combined profile in this direction gives 0.076 pc.  All these are very close to the FWHM we obtained without using \textit{getsf}. This result shows both that the sophisticated analysis in \textit{getsf} agrees with the straightforward analysis of Method 1 and that the value of the FWHM of the main filament is independent of the parameters that must be set to use \textit{getsf}.  
In the following analysis, we shall use the measurement of the FWHM by \textit{getsf}-- i.e., the actual FWHM, not a Plummer fit-- as the measure of the FWHM of subfilaments.

\subsection{Statistics of 2D physical properties of subfilaments of the main filament}
\label{sec:getsf}

\begin{table}
\caption{Median Physical Properties of Subfilaments and Segments Identified using 2D Data from \textit{getsf} for Different Projections, Beam Sizes and Values of $\wch$}
\label{tab:filament_prop_c}
\begin{tabular}{lccccccc}
\vspace{-0.3cm}\\
\hline
\hline
(1) & (2) & (3) & (4) & (5) & (6) & (7)\tablenotemark{d} & (8) \\
Time ($\tff$) & 0.3 & 0.3 & 0.5 & 0.5 & 0.5 & 0.5 & 0.5 \\
Projection\tablenotemark{a} & $y$ & $z$ & $y$ & $z$ & $z $& $z$ & $z$\\
$\theta_{\rm beam} {\rm (pc)}$ & 0.0066 & 0.0066 & 0.0066 & 0.0066 & 0.0066 & 0.0066 & 0.018 \\
$\wch {\rm (pc)}$ & 0.02 & 0.02 & 0.02 & 0.02 & 0.01 & 0.15 & 0.02\\
Number of subfilaments & 35 & 69 & 44 & 73 & 91 & 8 & 29\\
\hline
Length (pc) & 0.22 & 0.23 & 0.21 & 0.17 & 0.15 & 0.41 & 0.22\\
FWHM (pc) & 0.043 & 0.040 & 0.041 & 0.032 & 0.029 & 0.074 & 0.049\\
Aspect ratio & 4.6 & 5.3 & 4.6 & 6.2 & 5.5 & 6.8 & 5.8\\
Mass (\msun) & 4.4 & 2.0 & 7.9 & 2.4 & 1.19 & 31.0 & 3.5\\
Line mass (\msun\ pc\e) & 26.7 & 9.0 & 37.0 & 10.8 & 7.4 & 51.2  & 12.5\\
$ N_{\rm Hc} \tablenotemark{b} (\times 10^{22}\, {\rm cm}^{-2})$ & 1.6 & 0.74 & 1.1 & 1.1 & 0.86 & 4.6 & 1.1 \\
$ N_{\rm H,bg} \tablenotemark{c} (\times 10^{22}\, {\rm cm}^{-2})$ & 0.56 & 0.27 & 0.55 & 0.45 & 0.48 & 0.41 & 0.44\\
\hline
\hline
\end{tabular}
\tablenotetext{a}{The column density map in this projection
is used in subfilament identification.  The main filament is approximately oriented in the $x$ direction and the initial magnetic field is in the $z$ direction.}
\tablenotetext{b}{$ N_{\rm Hc}$ = central column density of a subfilament/segment above the background averaged along the subfilament/segment.}
\tablenotetext{c}{$ N_{\rm H,bg} $ = background column density from \textit{getsf} for subfilaments/segments.}
\tablenotetext{d}{Segments of the main filament; data in the other columns is for subfilaments.}
\end{table}
To show that subfilaments are distinct from the main filament and not an artifact of our choice of values for $\wch$, in Figure \ref{fig:fwhm_wch} we plot the median values of FWHM determined by \get\ as a function of the characteristic width, $\wch$, used in the analysis. We see that there are two regions in the plot: At $\wch=0.01-0.05$ pc, the FWHM is about 0.03-0.04 pc, corresponding to subfilaments, whereas for $\wch=0.125-0.2$ pc, the median FWHM is about 0.07-0.08 pc, corresponding to segments of the main filament.

As discussed above, subfilaments are closely related to the fibers detected by \citet{hacar13}, with similar widths, $\sim 0.02 - 0.05$ pc, and column densities, $\Nh \gtrsim 10^{22}$ cm\ee \citep[e.g.][]{hacar18, hacar23}.  In the simulation, subfilaments have nonthermal velocity dispersions comparable to the sound speed, so they are velocity coherent. These subfilaments are therefore essentially equivalent to fibers, with the exception that some observed fibers could be more than one subfilament that overlap in PPV space. Using numerical simulation, \citet{zamora2017} found that a minority by number but a majority by mass of velocity-coherent fibers identified using PPV data are really coherent in three dimensions.

The median values of the properties of subfilaments for $\wch=0.01$ and 0.02 pc and for segments of the main filament for $w_{\rm ch}= 0.15$ pc are listed in Table \ref{tab:filament_prop_c}.  We use the default setting in \textit{getsf} for the other parameters.

\begin{figure*}
\includegraphics[scale=0.54]{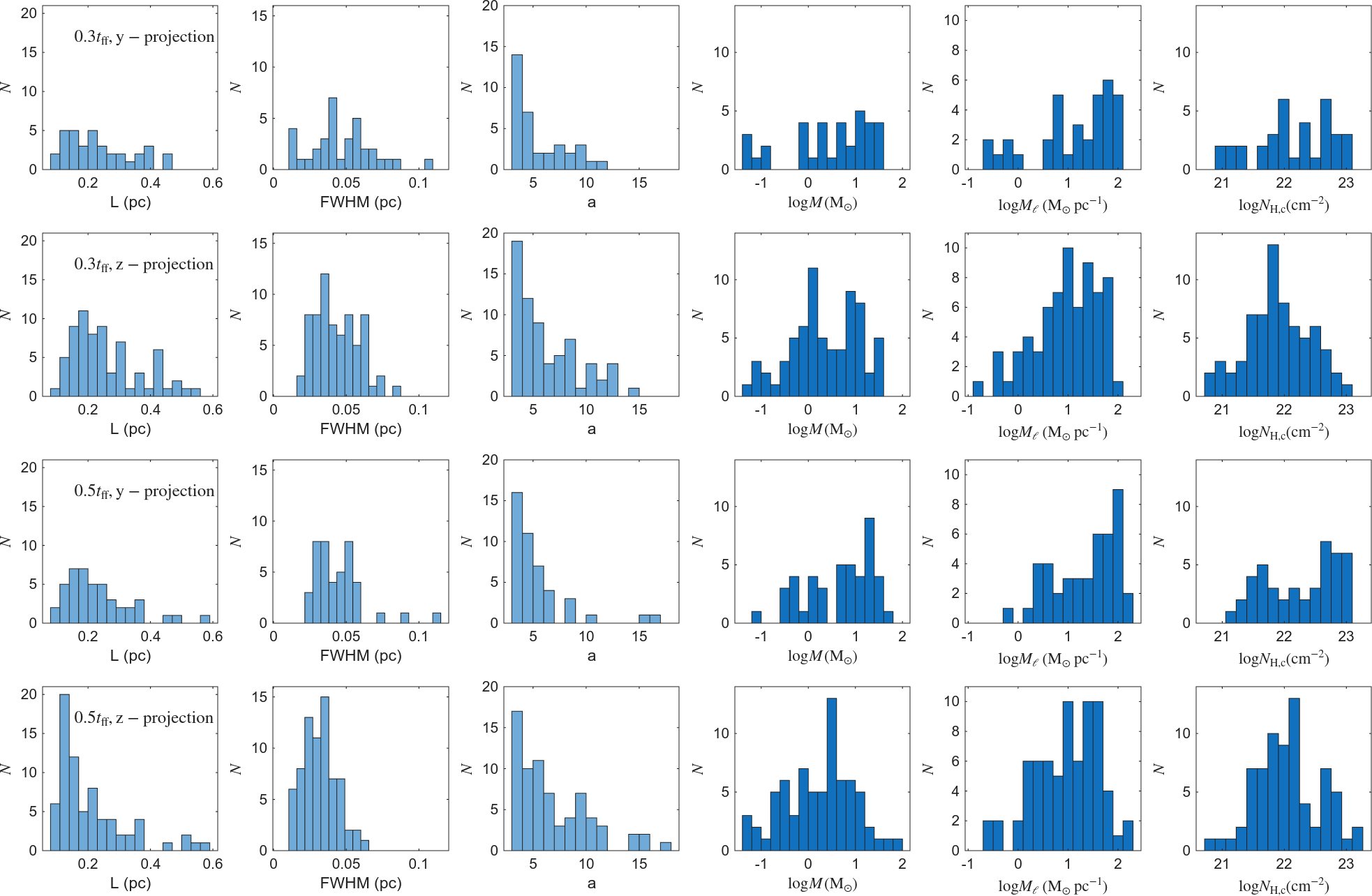}
\caption{Histograms of physical properties of subfilaments identifed by \textit{getsf} using $\wch = 0.02$ pc as the characteristic width for the background measurement.  The column density map is smoothed by a $\theta=0.0066$ pc beam.  The first row shows the histograms at time $0.3\tff$ by viewing along the y-axis.  The second row shows the histograms at time $0.3\tff$ but viewing along the z-axis.  The third and forth rows are the same except at time $0.5\tff$.  The columns of panels from left to right are length ($L$), FWHM, aspect ratio ($a = L/$FWHM), mass, ($M$), line mass, ($M_\ell$), and the central column density, ($\Nhc$).  The deeper color of the histograms in the last three columns is to highlight the fact that the horizontal axes are on a logarithmic scale.
\label{fig:hist_28as}}
\end{figure*}
In Figure \ref{fig:hist_28as}, we show the histograms of length, FWHM, aspect ratio (=length/FWHM), total mass, line mass, and mean column density of the subfilament structures of the 4 maps from two different times at two different projections (marked in the leftmost column of the figure panels) using $\wch = 0.02$ pc.  In Table \ref{tab:filament_prop_c}, we compare the effects caused by projection (columns (2) and (3), (4) and (5)), evolution of subfilaments (columns (2) and (4), (3) and (5)), the value of $\wch$ (columns (6) to (8)) and resolution due to beam size $\theta_{\rm beam}$ (columns (5) and (8)).  For example, the number of identified subfilaments is smaller when viewing the cloud along the $y$-axis than along the $z$-axis (35 vs 69) because the filament is flattened along the $z$-axis (the mean field direction).  As a result, subfilaments overlap more along the line-of-sight (los) when viewed along the $y$-axis.  
Table \ref{tab:filament_prop_c} shows that the median values of the FWHM at $0.3\tff$ are similar in different projections but the median is somewhat smaller in the $z$-projection at $0.5\tff$.  The median values of the FWHM for the segments (column 7) are close to the FWHM of the main filament, as expected. Resolution affects the median value of FWHM of filaments: Using $\theta_{\rm beam} = 0.018$ pc gives a median FWHM of 0.049 pc in the $z$ projection at $0.5\tff$, larger than the the value of 0.029 pc for $\theta_{\rm beam}=0.0066$ pc.  For $\wch=0.02$ pc, most subfilaments have a short length, $L < 0.3$ pc, and the median aspect ratios are less than twice the minimum.  The numbers of subfilaments in the two projections increase in time from $0.3 \tff$ to $0.5 \tff$.  Although subfilaments are merging in the main cloud, they are being created faster.  The mass and line mass of subfilaments are larger at $0.5\tff$ than at $0.3\tff$ as a result of the accretion of diffuse gas and mergers with other subfilaments, which are discussed in section \ref{subsec:filament_evol}.
Here we define ``merger" as an accretion process of more massive distinct gas objects such as two subfilaments or two cores.  It is different from diffuse gas accretion.  Over this period, the median line masses of the subfilaments increase somewhat, from $9-10.8$ \msun\ pc\e\ in the $z$ projection and $27-37$ \msun\ pc\e\ in the $y$ projection.  The peak column densities of the subfilaments are of order $10^{22}~ {\rm cm}^{-2}$ above the background level.

Observations show a power law relation of the mass and length of filaments, $L \propto M^\alpha$, with $\alpha \simeq 0.5$ \citep[e.g][]{hacar23, zhang2026}.  For the subfilaments identified by {\it getsf} in the simulation at $0.5 \tff$, we have $\alpha = 0.48\pm0.13$, consistent with the observed mass-length relation.

\begin{figure*}
\includegraphics[scale=0.54]{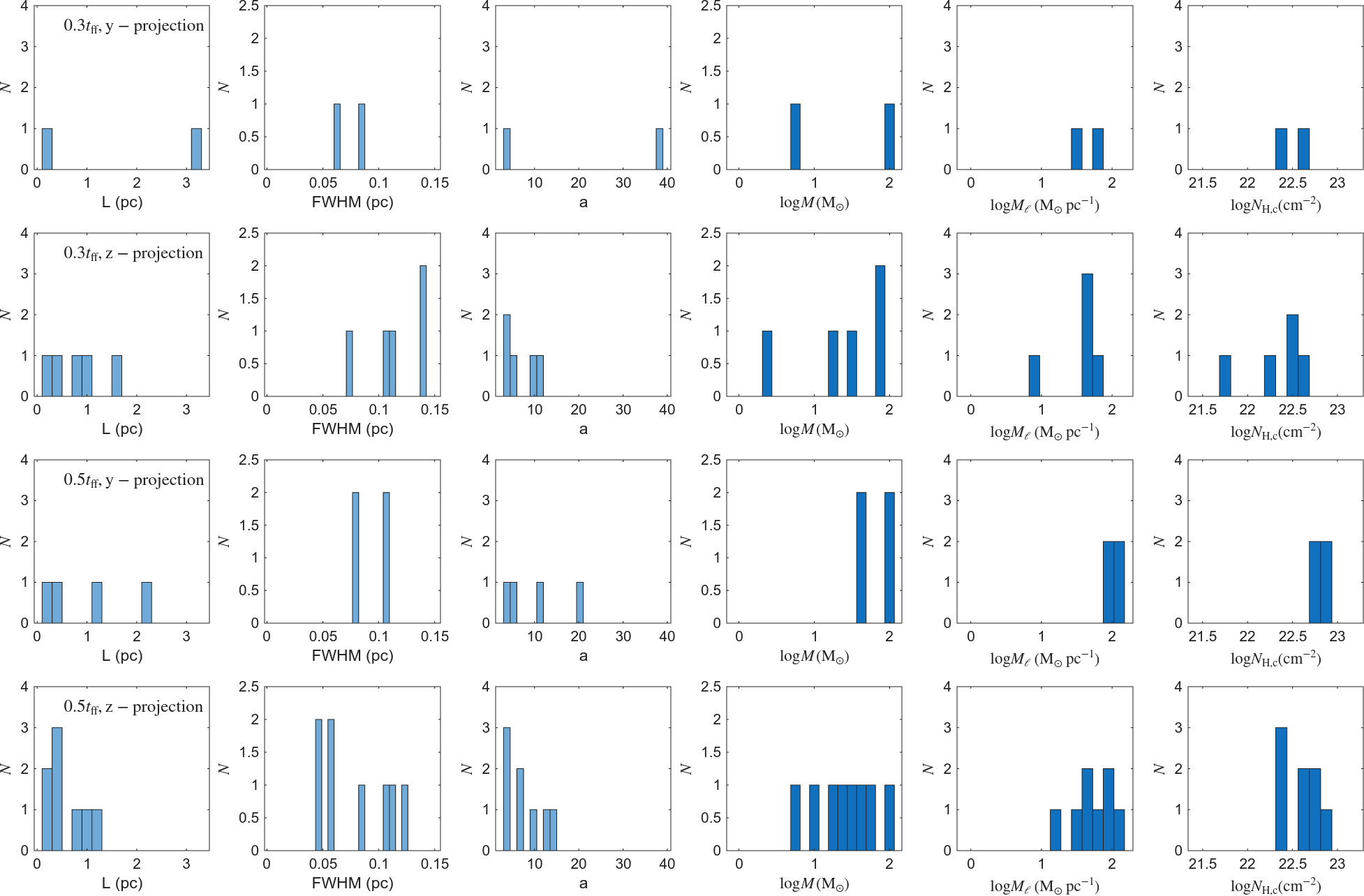}
\caption{Same as Figure \ref{fig:hist_28as} but using $\wch = 0.15$ pc for the background measurement.  With this choice of $\wch$, the focus is on segments of the main filament.
\label{fig:hist_221as}}
\end{figure*}

In Figure \ref{fig:hist_221as}, we show the physical properties of the segments of the main filament identified using $\wch=0.15$ pc.  The skeletons of the eight segments identified at $0.5\tff$ in the $z$-projection represent the main filament, as shown in the left panel of Figure \ref{fig:map+skeletons}.  In Figure \ref{fig:profiles}, we show the profiles of these segments. For the most part, they are well-fitted with Plummer profiles.  The median values of the physical properties of the main filament segments in Table \ref{tab:filament_prop_c} column (7) are larger than those of the subfilaments detected on smaller size scales.  The median values of the FWHM of the segments in column (7) are close to the FWHM of the combined profile of the main filament discussed in Section \ref{subsec:fwhm}.

\subsection{Evolution of subfilaments using 3D data}

\subsubsection{Formation
\label{subsec:filament_form}}

\begin{figure*}
\includegraphics[scale=0.27]{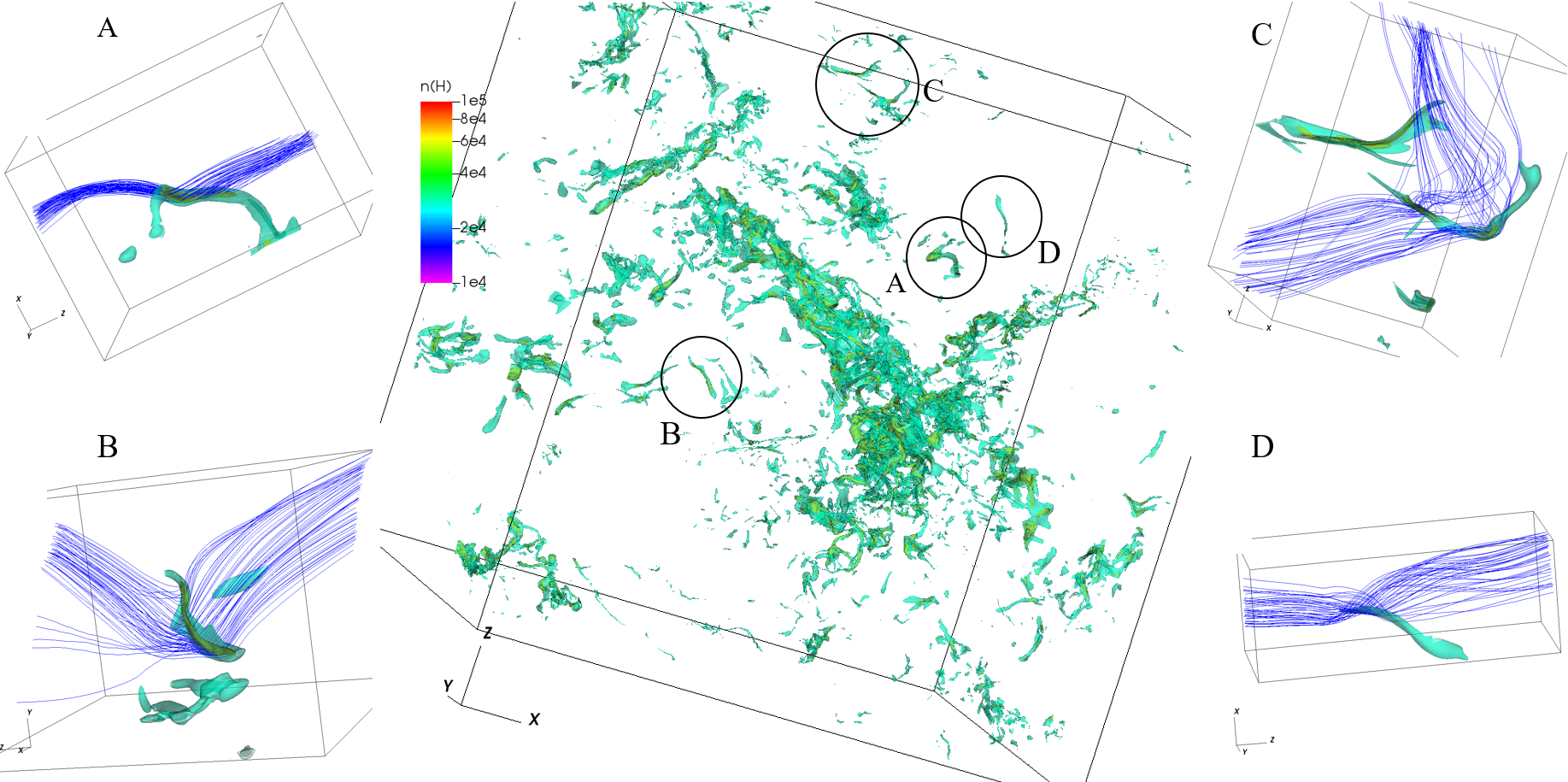}
\caption{The central part of the figure shows the isosurface density plot of the entire simulation region, focusing on the density range $\nh = 4\times 10^4$~cm\eee\
to $10^5$ cm$^{-3}$.  Four isolated filaments with magnetic fields are enlarged and plotted around the central image.
\label{fig:filament_Bfield}}
\end{figure*}
Instead of using projected 2D column density maps, from here on we analyze the formation and evolution of subfilaments using 3D data from the simulation.  As discussed in the Introduction, filaments can form via fragmentation of larger structures (``top-down") or by gas being swept together by shocks and/or converging flows (``bottom-up"; \citealp{hacar23,pineda23}). We find that the latter process is dominant in our simulation.  Here we discuss how shock compression and converging flows create filamentary structures in turbulent, magnetized, self-gravitating flows.

\begin{figure}
\begin{interactive}{animation}{A2.mp4}
\includegraphics[scale=0.2]{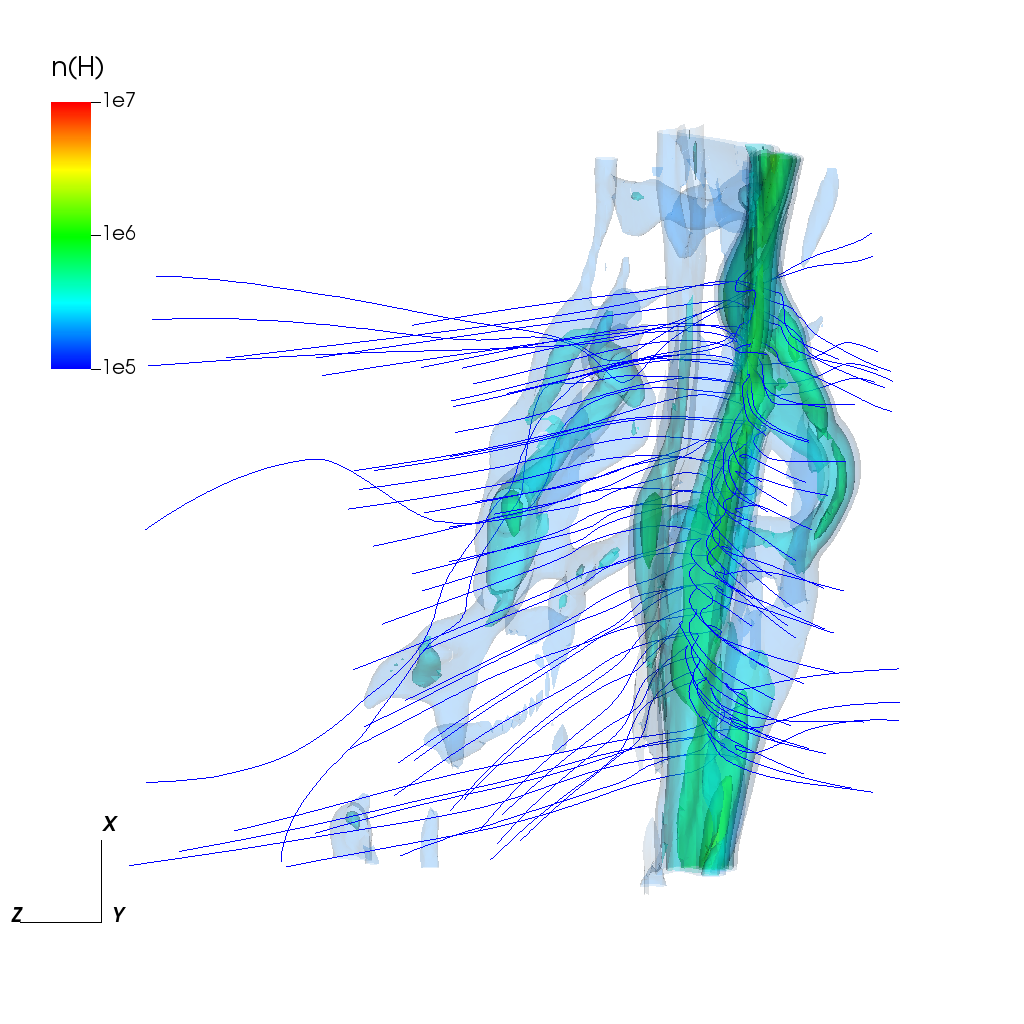}
\end{interactive}
\caption{Magnetic field structures of subfilaments at time 348 kyr inside a crowded region of the filament cloud in the simulation.  The animation A2 shows the evolution of the subfilaments and magnetic fields in this 0.27 pc region starting from 225 kyr to the end of the simulation.  \it{The animation associated with this figure is available in the online version of the published article on ApJ.}
\label{A2}}
\end{figure}

In Figure \ref{fig:filament_Bfield}, we show an isosurface plot of the gas density in the entire simulation region just before gravity is activated.  Four typical isolated filaments are enlarged to show the magnetic field through the filaments.  They are all located at kinks in the magnetic field, which are the result of oblique MHD shocks \citep{inoue2013,abe2021}.
These filaments are in an early stage of formation.  When these filaments move into the main filament and become subfilaments, they will have more interactions with each other.  The kinks of the magnetic field associated with some of the more massive subfilaments inside the cloud could be the result of the magnetic field being dragged by the movement of subfilament through turbulent flows.  We can see more of this field structure in the animations A2 and A3 (Figures \ref{A2} and \ref{fig:helicalfield}).

\begin{figure*}
\includegraphics[scale=0.33]{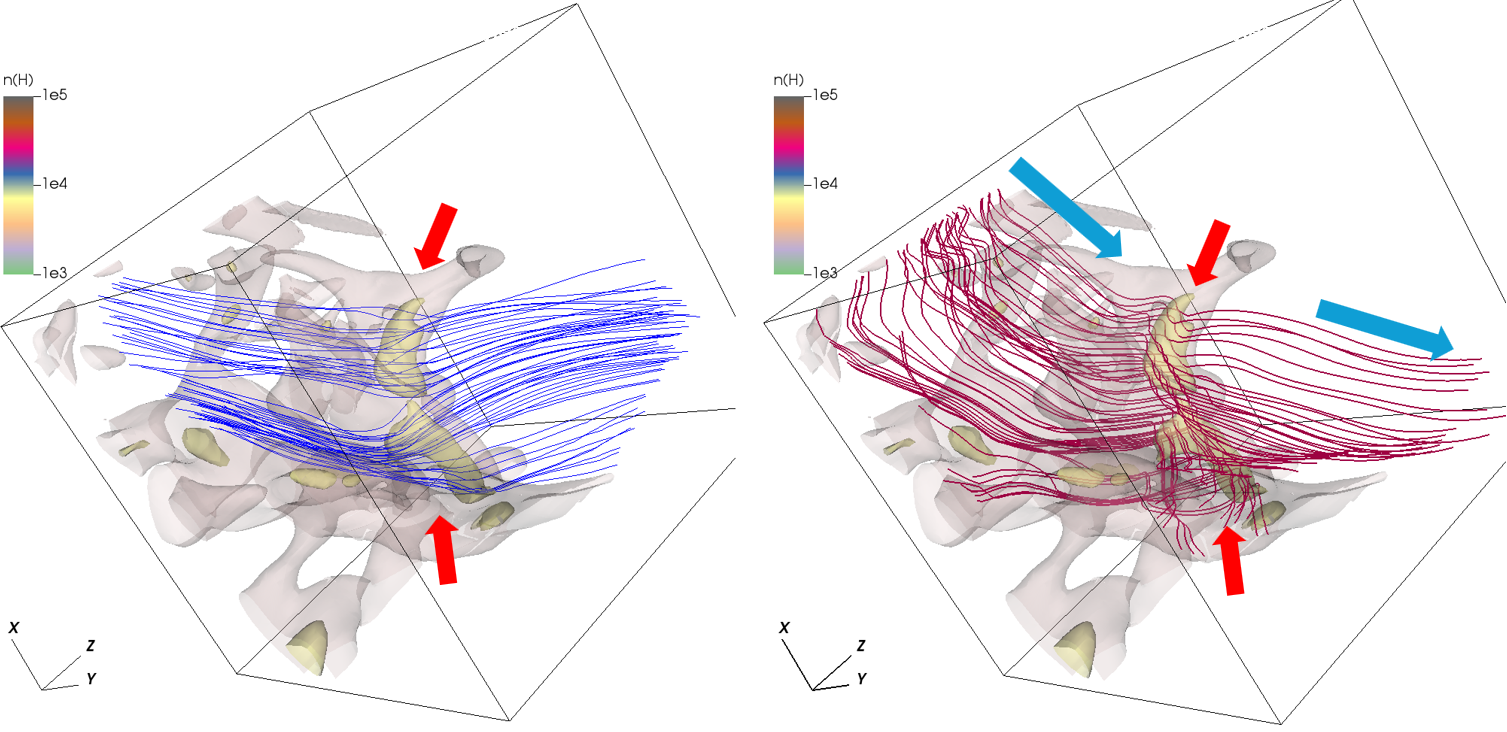}
\caption{Left: Isosurface plot of density around elongated gas structures showing the magnetic field lines in a 0.27 pc region, focusing on the density range $\nh = 1\times 10^3$~cm\eee\ to $10^4$ cm$^{-3}$.  The elongated gas structures marked by the red arrows have just started to form and are located at the kinks of the magnetic field.  The volume density $\nh \la 10^4$ cm$^{-3}$ at this moment.  Right: Same as the left panel but velocity streamlines are plotted.  Blue arrows indicate the gas flow direction.  The velocity field lines have an abrupt change in direction at the two elongated gas structures, indicating the shock locations.
\label{fig:BVstreamlines}}
\end{figure*}

In Figure \ref{fig:BVstreamlines}, we show the very early stage of the formation of a shock-compressed filamentary structure.  In Figure \ref{fig:BVstreamlines}a, the streamlines of the magnetic field are plotted with the isosurface density of the gas structure in this region, which is 0.27 pc in length.  The two dense elongated gas structures marked by the red arrows are the result of gas flowing along the field lines and accumulating at the kinks, which are the shock locations.  The density of the elongated gas structure is $\nh\sim 10^4$ cm$^{-3}$.  The velocity streamlines are plotted in \ref{fig:BVstreamlines}b.  At the shock locations marked by the red arrows, the velocity streamlines abruptly change direction. Gas flows in the direction indicated by the blue arrows and accumulates at the shock; later the two elongated structures connect and become a filament, with an aspect ratio $>3$.  Because the gas flows along the magnetic field, it will be perpendicular to the long axis of the filament. The final mass of the filament formed by turbulent compression depends on how long the converging flows persist, which results in a range of line masses for the filaments.  In Section \ref{subsec:filament_evol}, we shall see that converging flows lead to filament mergers.

\subsubsection{The role of subfilament mergers}
\label{subsec:filament_evol}

\begin{figure*}
\includegraphics[scale=0.9]{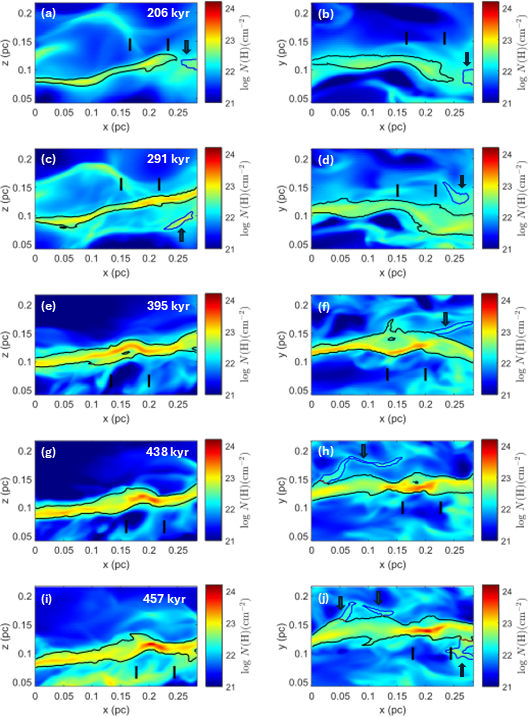}
\caption{Subfilament collision and merger with nearby subfilaments and gas structures based on 3D data.  The left and right figure panels are maps of the column densities, $\Nhy$ and $\Nhz$, along the $y$ and $z$ axes, respectively.  The times of the images in each row after gravity is activated are marked on the left column panels.  Black curves show the projected edges of the subfilament identified by CLUMPFIND; the physical properties of this subfilament are listed in Table \ref{tab:filament_prop_a}.  The two short lines indicate the portion (0.07 pc) of the subfilament that has its $\ml / M_{\ell,{\rm crit}}$ values shown in Table \ref{tab:filament_prop_a}.  The blue contours indicated by the black arrows are nearby gas structures in the process of collision and merging with the subfilament.  See Section \ref{subsec:filament_evol} for discussion on the filament stability.
\label{fig:filament_merge}}
\end{figure*}

As seen in Paper II and Figure \ref{fig:filament_cloud} above, there is a large number of sub-parsec long, low density filaments  near the converging locations of large-scale turbulent flows.  With the help of turbulent converging flows, these filaments come together, collide, and merge to form denser subfilaments in the main filament due to gravity.  Without gravity, these filaments would collide but could not form massive filaments since turbulence tends to tear them apart \citep{li19}.

In Figure \ref{fig:filament_merge}, we show the evolution of a subfilament as it accretes and merges with nearby subfilaments, resulting in an uneven increase in line mass.  This subfilament was randomly selected from among the subfilaments that form a single core during the monitoring period.  We used the CLUMPFIND algorithm \citep{williams1994} to determine the 3D structure of the subfilaments in our simulation; note that \get\ supplies only 2D information.  CLUMPFIND requires a minimum density to operate. We note that the typical density in gas shocked at the mean Mach number of 10 in our simulation is $100\nbh\simeq 1.0\times 10^5$~cm\eee.  We have found that setting the threshold density at several times this, $\nthr=3\times 10^5$~cm\eee, enabled us to distinguish different subfilaments in the complex ambient medium in the simulation.  This threshold defines the surface of the subfilaments in 3D space so that we can compute the physical properties of the subfilaments.  The five rows of the figure panels in Figure \ref{fig:filament_merge}  correspond to five time snapshots of the selected subfilament projected along the $y$- and $z$-axes in the left and right columns, respectively; the subfilament is within 12 degrees of the $x$-direction.  The time interval extends from just before the collision, $t=206$ kyr after gravity is activated, to when the dense part of the subfilament has reached the critical line mass at $t=457$ kyr.  The average line mass of the subfilament in the 0.28 pc window, $\ml$, the geometrical mean column number densities, $\langle \Nh \rangle$, obtained from the projected column densities in the y- and z-projection, the root-mean-square magnetic field strength, $B_{\rm rms}$, and the 1D velocity dispersion, $\sigma$, perpendicular to the axis of a long subfilament, are computed and listed in Table \ref{tab:filament_prop_a}.  The ratios of the line mass to the critical value are integrated for all connected cells with volume density greater than $3\times10^5$ cm$^{-3}$.
\begin{table}
\caption{Time Evolution of Physical Properties of a Subfilament (Figure \ref{fig:filament_merge}) Using 3D Data from the Simulation\tablenotemark{a}}
\label{tab:filament_prop_a}
\begin{tabular}{lcccccc}
\vspace{-0.3cm}\\
\hline
\hline
Time (kyr) & 206 & 291 & 395 & 438 & 457 \\
\hline
Line mass $\ml$ ($M_\odot~{\rm pc^{-1}}$) & 4.76 & 8.88 & 17.4 & 18.4 & 19.4 \\
1D velocity dispersion $\sigma$ (km s$^{-1}$) & 0.23 & 0.23 & 0.26 & 0.25 & 0.24 \\
Mean column density $\langle N_{{\rm H}} \rangle$ ($\times 10^{22}$cm$^{-2}$)\tablenotemark{b} & 1.50 & 2.31 & 4.00 & 4.79 & 4.08\\
Magnetic field strength $B_{\rm rms}$ ($\mu G$) & 91.6 & 121 & 190 & 154 & 132 \\
$\ml / M_{\ell,{\rm crit}}\tablenotemark{c}$ & 0.19 & 0.32 & 0.46 & 0.57 & 0.59 \\
\hline
$\ml / M_{\ell,{\rm crit}}$\tablenotemark{c,d} & 0.27 & 0.44 & 0.53 & 0.83 & 1.05 \\
\hline
\end{tabular}
\tablenotetext{a} {All quantities in the table are based on cells identified by CLUMPFIND with $\nh>3\times 10^5$ cm\eee\, and are averaged over the window of length 0.28 pc shown in Figure \ref{fig:filament_merge}.}
\tablenotetext{b} {The geometrical mean of the column densities in the y- and z-projections, averaged over the mean projected width of the subfilament.}
\tablenotetext{c} {$M_{\ell,{\rm crit}}$ is evaluated from Equation (\ref{eq:mellcr}). The mass and flux in $\mu_\Phi$ are based on the geometric mean of the projected widths of the subfilaments.}
\tablenotetext{d} {Evaluated for the short section of 0.07 pc marked in Figure \ref{fig:filament_merge}; a core forms there.}
\end{table}

In Figure \ref{fig:filament_merge}, the black curves show the projected edges of the subfilament identified by CLUMPFIND.  The blue contours, pointed to by black arrows, are the nearby subfilaments and clumpy gas structures that will collide and merge with the subfilament during this time interval.  As a result, the line mass and magnetic field strength increase in time.  The subfilament changes from magnetically subcritical to supercritical.  A dense region on the right is marked by two short vertical lines.  This part of the subfilament eventually evolves into a dense core with mean density $\nh > 10^6$ cm$^{-3}$.  In the top row of the figure, we see is very close to that of a subfilament on the right indicated by an arrow. These two subfilaments collide and merge into one, as shown in Figure \ref{fig:filament_merge}c and d.  The subfilament later collides and merges with a gas clump and another less massive subfilament at the location indicated by the arrows in Figure \ref{fig:filament_merge}c, d, and f.  The subfilament gains a significant amount of mass and turbulent energy in that region.  Subsequently, a dense core forms there as shown in Figure \ref{fig:filament_merge}g-j. 
We computed the physical values of this short section (0.07 pc) and list just the $\ml / M_{\ell,{\rm crit}}$ in the last row of Table \ref{tab:filament_prop_a}.  A part of a low density subfilament, indicated by an arrow in Figure  \ref{fig:filament_merge}h, merged with the left part of the subfilament under consideration but was not massive enough to create an unstable condition at the end of this analysis at 457 kyr.

For the entire length of the subfilament in the 0.28 pc long window, the line mass began at $M_\ell\simeq 5$~\msun~pc\e\ with a FWHM of $W_{\rm P}\simeq 0.025$ pc and grew to $M_\ell\simeq 19$~\msun~pc\e\ with $W_{\rm P}\simeq 0.016$ pc over a period of 250 kyr. At the same time, the central density increased from $0.9\times10^6$ cm\eee\ to $5.4\times10^6$ cm\eee.  The average of the central column density to the mean column density, $\NHc/\langle\NH\rangle$, of the subfilament is about 3.1.  This subfilament is therefore similar to the ``fibers" observed in Orion by \citet{hacar18}.

This subfilament is supported against gravity by thermal pressure, turbulent pressure, and a perpendicular magnetic field; the axial and azimuthal magnetic fields are weaker than $B_{\rm rms}$ and do not contribute significantly to the forces in the subfilament.  The mean values of axial and azimuthal components are $17.9 \sim 21.3 \mu$G and $1.7 \sim 4.4 \mu$G, respectively.  We compute the axial and azimuthal components by vector summation of the $B$-field of all the computational cells inside the subfilament surface along and around the spine of the subfilament.  The ratio of line mass to the critical value (Eq. \ref{eq:mellcr}) changed from $\ml/M_{\crit,\ell} = 0.19$ at 206 kyr to 0.57 at 457 kyr.  The fact that $\ml/M_{\crit,\ell} < 1$ indicates that, on average, the subfilament can be in hydrostatic equilibrium.  Note that the support of this subfilament in this period of time is dominated by kinetic energy ($\calm_{\rm A}^2> 1$ based on the results in Table \ref{tab:filament_prop_a}).  The line mass in the 0.07 pc long marked region is up to 1.6 times greater than the average of the subfilament.  By 457 kyr, the line mass has reached the critical value  as a result of collision with another subfilament (see the discussion in Section \ref{subsec:filament_evol}), and much of this part of the subfilament is actually in the embedded core.
\begin{figure*}
\includegraphics[scale=0.53]{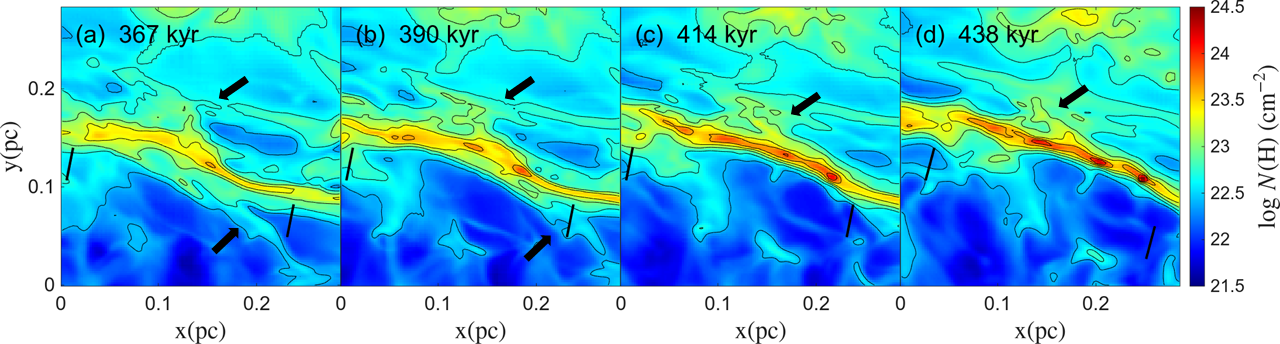}
\caption{Column density map of another unstable subfilament as the result of merger with nearby subfilaments, marked by black arrows, at times (a) 367 kyr, (b) 390 kyr, (c)  414 kyr, and (d) 438 kyr.  The values of $M_\ell / M_{\ell,{\rm crit}}$ of the region between the two black-colored lines are 0.75, 0.89, 1.03, and 1.17, respectively, from panel (a) to (d). The column density contours are log $\Nh$=[22.4,22.7,23.0,23.3,23.6,23.9] cm$^{-2}$.  See Section \ref{subsec:filament_evol} for discussion.
\label{fig:unstable_filaments}}
\end{figure*}

In Figure \ref{fig:unstable_filaments}, we show another example of an unstable subfilament as a result of the merger of subfilaments.  
The subfilament of about 0.23 pc long, the region between the two black-colored lines, is shown at four different times from 367 to 438 kyr.  At 367 kyr, the subfilament has been merging with subfilaments from above in the column density maps. The mass of the subfilament increases from 21.5 to 38.9 \msun\ during this period of time shown in Figure \ref{fig:unstable_filaments}.  The values of $\ml/M_{\crit,\ell}$ of the subfilament are 0.75, 0.89, 1.03, and 1.17, respectively at these four times.  At 414 kyr, the subfilament is theoretically unstable 
and cores begin to form.
\begin{figure*}
\begin{interactive}{animation}{A3.mp4}
\includegraphics[scale=0.68]{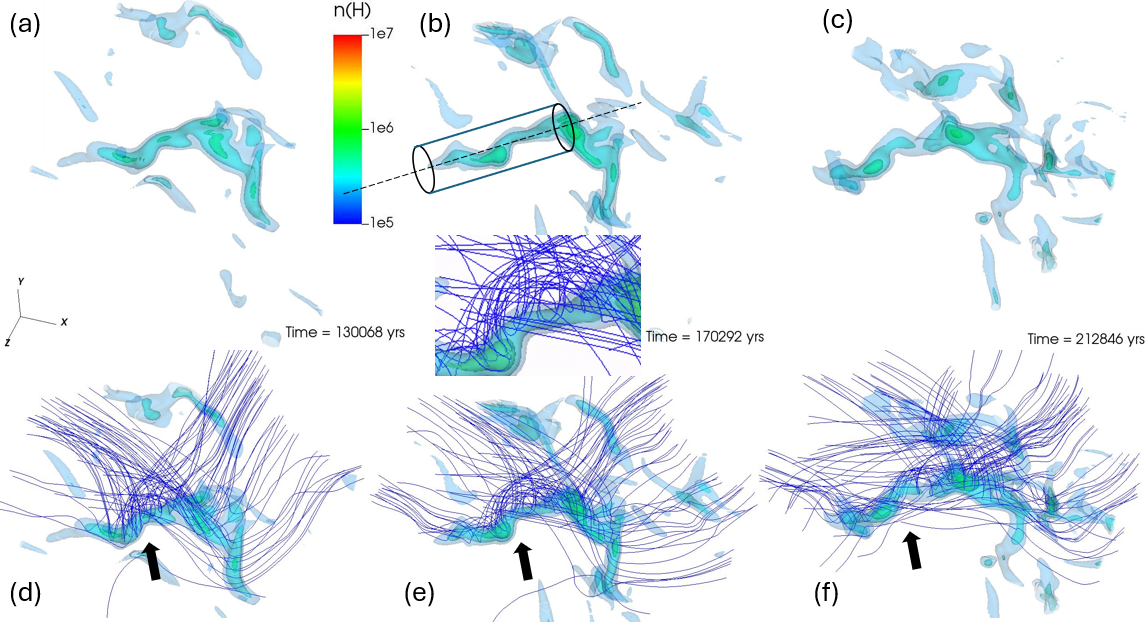}
\end{interactive}
\caption{Evolution of a subfilament with the development of a helical magnetic field.  (a) - (c) are the volume rendering of the density at three different times.  (d) - (f) show the magnetic field structures at the corresponding times.  The black arrows point to where a helical field has developed at $t=170$ kyr and the subfilament is cut in two parts $t=213$ kyr.  The cylinder shown in (b) is the space where the values of magnetic field components are computed for an unstable subfilament discussed in Section \ref{subsec:helicalfield}.  The inset between panels (b) and (e) is a magnified portion of the subfilament in panel (e) to visualize the helical field.  The color bar for the density applies to all panels.  This period of time shown in animation A3 is before the formation of Core 1 discussed in Section \ref{subsec:core1}.  \it{The animation associated with this figure is available in the online version of the published article on ApJ.}
\label{fig:helicalfield}}
\end{figure*}

\subsection{Example of the effects of a helical magnetic field}
\label{subsec:helicalfield}

\citet{fiege00a} considered the case of a filament with a helical magnetic field. They pointed out that the toroidal field actually helps gravity squeeze the filament.  It is not possible to follow the evolution of the hundreds of subfilaments in the simulation, so we followed the time history of about 30 subfilaments and found one case in which a helical field persisted for more than 40 kyr.  In the top row of Figure \ref{fig:helicalfield}, we show a volume rendering of the density at three different times.  The bottom row of panels shows the same density rendering, but with the addition of magnetic-field lines.  The black arrow points to where the helical field develops in time.  The subfilament is well developed at 130 kyr, but the kink-shaped magnetic field is not yet in a helical structure.  The gas accumulates at the kink and forms a subfilament.  A helical magnetic field with significant toroidal and poloidal components begins building up at about 130 kyr, and at 170 kyr the magnetic field at the marked location has a fully developed helical structure.  At this time, the mean values of the magnetic field components inside a cylindrical volume of length 0.102 pc and diameter of 0.026 pc (see Figure \ref{fig:helicalfield}b) around the subfilament at the location of the black arrow in the figure are $B_\parallel = 47\pm5 \mu$G, $B_\phi=50\pm3 \mu$G, and $B_\varpi=30\pm1 \mu$G.  The cylinder is constructed with the axis goes through the two end points on the spine of the serpentine-like subfilament.  The uncertainties of the three components are obtained by varying the diameter of the cylindrical space from 0.026 pc to 0.013 pc.  Compared to the $B_\phi < 10 \mu$G of the subfilament shown in Figure \ref{fig:filament_merge}, the $B_\phi$ in this case is significantly larger and comparable to $B_\parallel$.  Around 213 kyr, the subfilament breaks into two clumps--the helical field acts like a scissors cutting the subfilament into two parts.  Part of the subfilament still has a helical field.  The helical field has dissipated by $t\simeq 220$ kyr.  Helical fields appeared at several other locations in the simulation where shearing flows are strong, but they did not last as long as this one.
If the subfilament breaks due to the helical field before it becomes supercritical, the pieces of the subfilament diffuse back into the turbulent flow.  If the subfilament is close to supercritical, the fragmentation induced by the helical field may result in the formation of dense cores.  In the example shown in Figure \ref{fig:helicalfield}, the subfilament breaks into two parts (on the two sides of the black arrow).  The gas on the right-hand side evolves into a core that will become Core 1, which we shall discuss in Section \ref{subsec:core1}.

\section{Core Formation and Evolution: General}
\label{sec:core}

\citet{andre14} define a dense core as ``an individual fragment or local overdensity which corresponds to a local minimum in the gravitational potential of a molecular cloud." Observers do not have access to the density and so must rely on the column density to define cores. Since we have access to the density in our simulation, we define a dense core as a gas fragment with (1) a density at the surface greater than a threshold value, $\nthr$ and (2) a central density at least twice the ambient density.  The ambient density is defined as the mean density of the gas with $\nh<\nthr$ and inside the 0.15 pc size box where the core is located. As discussed in Section \ref{subsec:filament_evol}, we set the threshold density at $\nthr=3\times 10^5$ cm\eee. To ensure that dense cores differ from subfilaments, we require that dense cores have aspect ratios less than 2.  To ensure that cores are at least moderately resolved, we require a minimum equivalent spherical diameter $> 0.009$ pc (4 cells), corresponding to at least 34 computational cells.  When dense cores become gravitationally bound, they become pre-stellar cores. We adopt the criterion that a core is bound, at least approximately, when the virial parameter satisfies $\avir<2.5$ based on the discussion of Equation (\ref{eq:avir2}) below.

\subsection{Core formation by subfilament collisions}

The analysis in Section \ref{subsec:filament_evol} and animations A1 and A2 show that the main filament is full of subfilaments interacting with each other, with frequent supersonic and subsonic collisions.  Because these collisions are inelastic, the colliding subfilaments often merge.  Mergers result in an increase of mass and kinetic energy, and often magnetic energy, depending on the orientations of the subfilaments.  The merger of two subcritical subfilaments can produce a supercritical subfilament, provided the initial subfilaments are sufficiently magnetically supercritical. If the subfilaments collide at an angle, the contact location could become unstable and a core could form, as seen in Figure \ref{fig:filament_merge}.  In one case, we observe three pre-stellar cores forming at 438 kyr (Figure \ref{fig:unstable_filaments}) close to each other along a massive subfilament, forming a chain of cores like those observed in L1495/B213 by \citet{tafalla2015}.

About 650 cores have formed by the end of the simulation. Of these, 47 are at least of moderate mass ($M>1$~\msun).  We randomly chose nine of them for examination.  Most of these nine cores were found to have lifetimes $>500$ kyr.  Two of these cores are analyzed in greater detail in Section \ref{sec:two}.  These nine cores are the result of subfilament collisions and subsequent mergers.  While this is only a small fraction of total 652 cores of different masses formed in the simulation (see Section \ref{sec:cmf}), it is clear that collisions and mergers of subfilaments are important in core formation. This is quite different from the standard picture in which cores are formed via gravitational instability in shocked layers of gas, rather than in filaments \citep[e.g.][]{whitworth1994,padoan2011}.  Because when, where, and how subfilaments collide inside turbulent molecular clouds are determined by turbulence, predictions of the spacing of dense cores using linear perturbation theories are unreliable.  We note also that theoretical estimates of the spacing of the cores (e.g., \citealp{nagasawa87}) are based on the assumptions that (1) the filaments have a column density distribution given by Equation (\ref{eq:plummer}) with $p=4$, whereas both observed and simulated filaments typically have $p\simeq 2$ and (2) that the field is parallel to the filament, whereas in our case the field is mainly perpendicular to the filament.

\subsection{Growth of Dense Cores by Mergers}
\begin{figure*}
\includegraphics[scale=0.63]{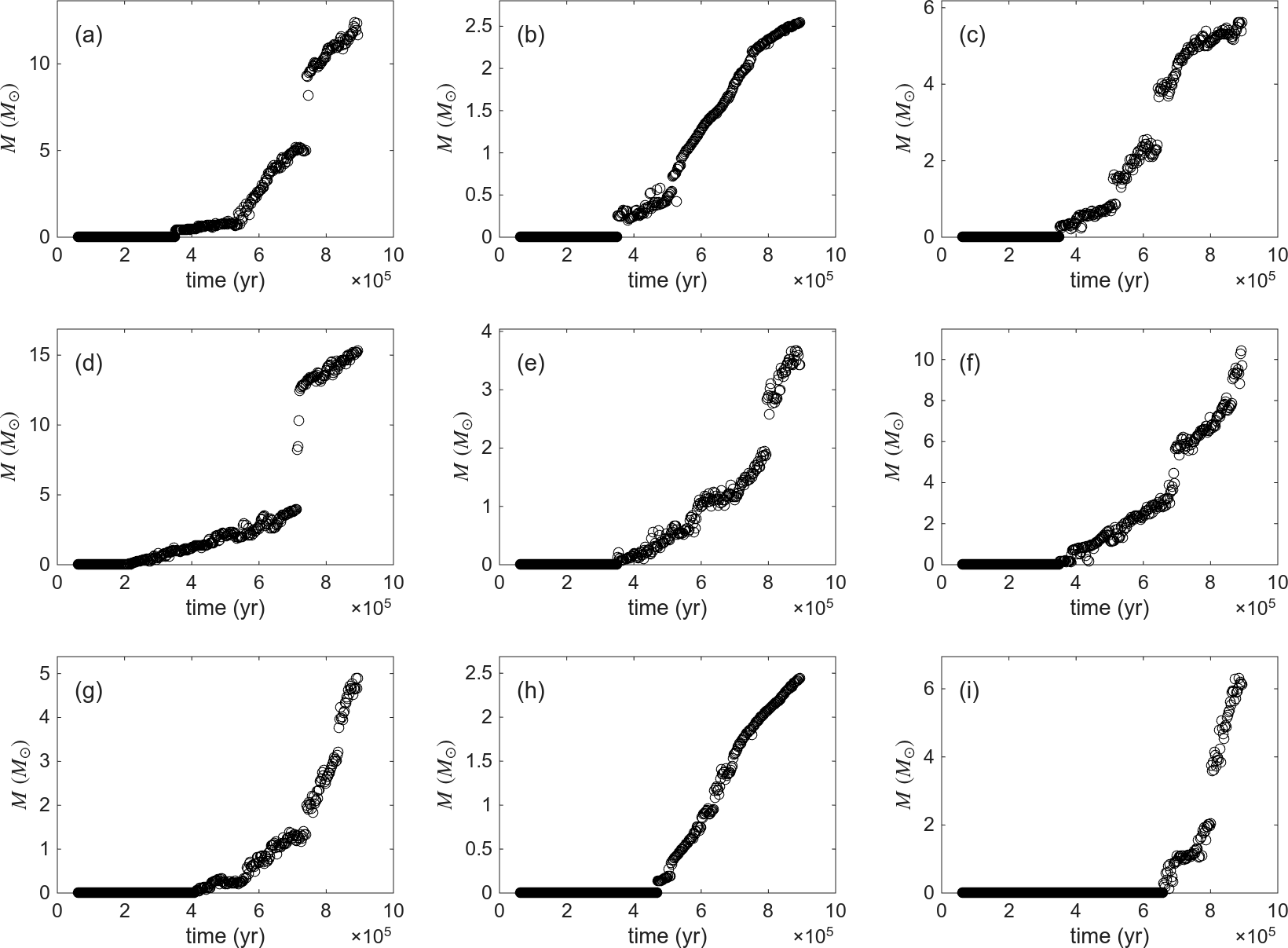}
\caption{Histories of mass accretion of nine cores.  The two cores at the top left are Core 1 and Core 2, which are also shown in Figure \ref{fig:core1p}a and \ref{fig:core2p}a.  The median terminal core mass is 5.6 \msun.
\label{fig:coremass}}
\end{figure*}

We find that core mergers play a significant role in the growth of cores.  The mass accretion histories of the nine cores selected in the previous section are plotted in Figure \ref{fig:coremass}.  The time evolution of a core in a simulation is nontrivial to trace. The numerical method used for this tracking is described in Appendix A.  At the end of the simulation, the nine cores have masses of 2.5 to 15 \msun.  The median value of the time-averaged accretion rate (including mergers) of the nine cores is $1.0\times 10^{-5}$~\msun\ yr\e, with a dispersion of 0.4 dex.
The time-averaged mass accretion rate for a core is defined as $\Delta m / \Delta t$, where $\Delta m$ is the total mass change from the time when the core can be identified to the end of the simulation, and $\Delta t$ is the corresponding time interval; $\Delta t$ ranges from $230 - 677$ kyr.  The typical accretion time scale for the cores exceeds $5\times 10^5$ yr and was determined by the run time of the simulation.  As said in Appendix A, the cores are tracked backward in time until the density of the densest cell is comparable to the nearby gas.  That is the beginning of the lifetime of a core.  We define a major merger as one in which there is an increase in mass of at least a factor 1.5 between consecutive data points, and a significant merger as one in which the increase is a factor of 1.2. Four of the nine cores (Figure \ref{fig:coremass}a, c, d, and i) experienced a median number of two major mergers; the median value of the final mass of these cores is $(9.3 \pm 3.4)$ \msun, whereas the corresponding value for the other five cores is $(3.7 \pm 1.2)$ \msun.  The median accretion rates of the four cores with major mergers and the other five cores are $2.3\times10^{-5}$ \msun\ yr\e and $1.03\times10^{-5}$ \msun\ yr\e, respectively.  All nine cores experienced at least two, and a median number of five, significant mergers. Two of the nine cores obtained more than 50\% of their terminal mass and five of the nine cores obtained more than 40\% of their terminal mass from significant mergers. The core shown in Figure \ref{fig:coremass}b, which is the core discussed in Section \ref{subsec:core2}, and in Figure \ref{fig:coremass}h are examples of cores that are relatively isolated from other cores and do not experience major mergers.  

We note that core mergers could play a role in the formation of low-mass binaries, adding to the many theories for binary formation \citep{tohline02}. Core mergers can bring the dense inner regions of the cores that will form protostars into close proximity, thereby creating the possibility of the formation of a binary star in a very dense environment. Such a mechanism would have to be investigated at much higher resolution than our simulation.

\subsection{The Core Mass Function}
\label{sec:cmf}

\begin{figure*}
\includegraphics[scale=0.35]{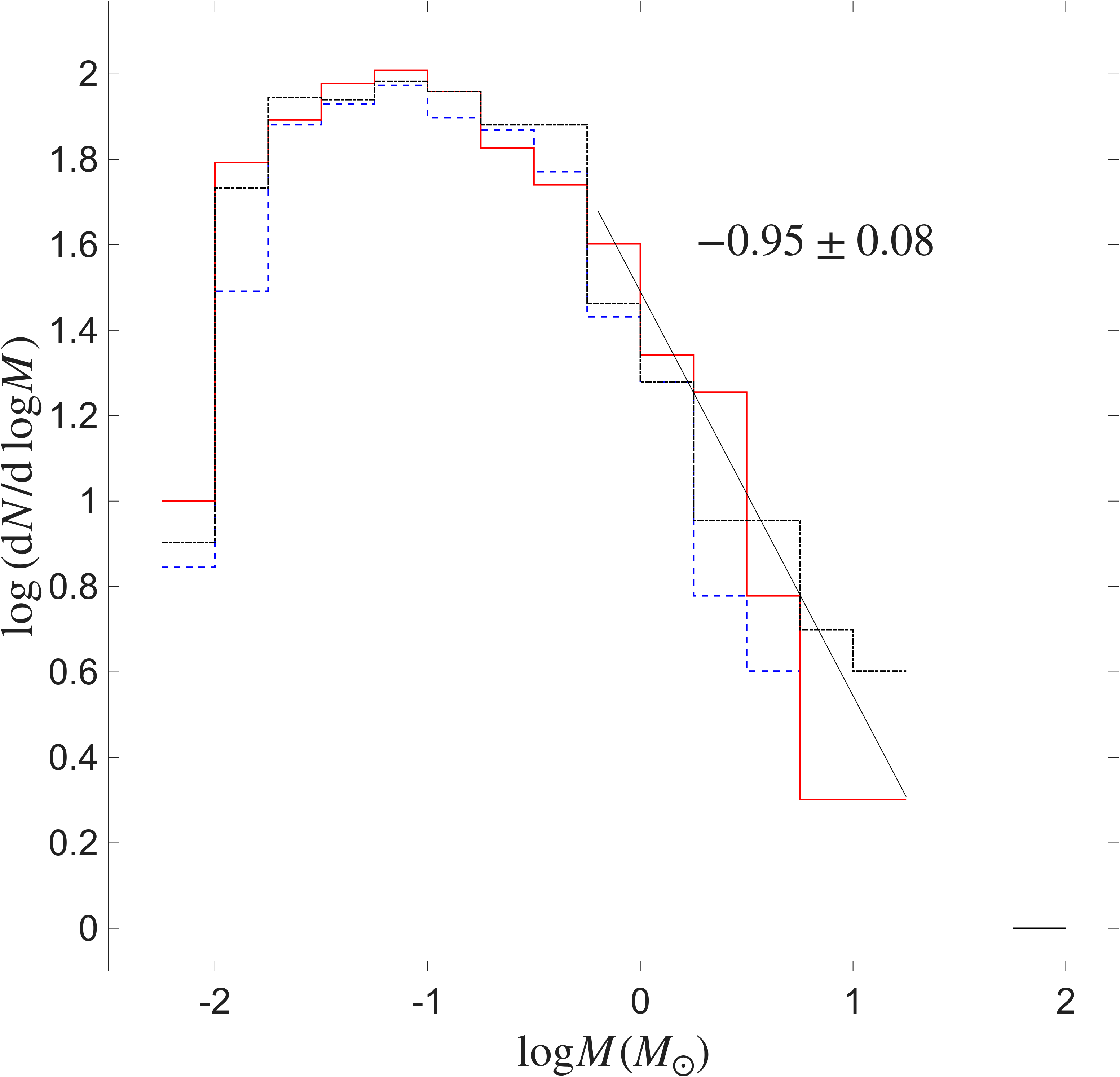}
\caption{Core mass functions at times 658 kyr (black dot-dash), 776 kyr (red solid), and 894 kyr (blue dash).  For clarity, we show the CMF at only three times in the figure.  The black line is the mean power law of the best-fit core mass functions on five data sets from 658 kyr to 894 kyr; the power-law index is shown in the figure.  The peak of the function is about 0.075 \msun\ in this time interval.
\label{fig:core_cmf}}
\end{figure*}

In Figure \ref{fig:core_cmf}, we show the core mass functions (CMFs) of the cores defined above in the entire simulation region at times of 658, 776, and 894 kyr.  There are 565 cores of total mass 185 \msun\ identified at a time of 658 kyr.  The maximum core mass is 15 \msun.  At 894 kyr, there are 652 cores with a total mass of 347 \msun and a maximum core mass of 78 \msun. On average, 75\% of the cores are gravitationally bound, with $\avir<2.5$. (This approximate criterion is larger than the standard value of 2 to allow for a centrally concentrated density--see the discussion below Eq. \ref{eq:avir2}).  The peaks of the mass function at these two times are both 0.075 \msun.  This is comparable to the peak of \citet{kroupa01}'s stellar IMF at 0.08 \msun, whereas one expects the CMF to peak at a mass several times greater than the stellar IMF \citep{mckee07}. The origin of this discrepancy is not clear, but we note that our definition of a core based on a density threshold is not the same as the definitions used by observers.  \citet{morii2026} determined the CMF from 715 cores detected in the ALMA Survey of 70mm Dark High-mass Clumps in Early Stages(ASHES). The resulting CMF peaks at about 0.2 \msun, but this value is uncertain because the sample is 90\% complete only above 1.2 \msun. Based on Herschel survey, the CMF obtained by \citet{benedettini2018} has the peak between 0.1 and 0.2 \msun, with completeness of 90\% to 0.1 \msun.  The CMF stops at about $7.5\times10^{-3}$ \msun is because of the our requirements of core at the outset of Section \ref{sec:core} must have at least 34 cells and a minimum density of $3\times 10^5$ cm\eee, which implies a minimum mass of $4\times10^{-3}$ \msun.  For clarity, we show the CMF at only three times in Figure \ref{fig:core_cmf}.  The mean power-law index at the high-mass end of the CMF, starting from the bin of 0.75-1 \msun and obtained from five data sets equally spaced in time from 658 to 894 kyr, is $-0.95\pm0.08$, consistent with the observed CMFs \citep[e.g.][]{fiorellino2021,louvet2024,morii2026}.

\section{CORE FORMATION AND EVOLUTION: TWO CORES}
\label{sec:two}

Every core has its own story of evolution.  In the following, we give two examples of core formation and evolution to show the physical changes in the cores from the time they are formed to the end of the simulation.  Core 1 is subjected to several mergers with gas subfilaments and nearby cores in a crowded environment, whereas Core 2 forms in a less crowded region and has only one minor merger.

\subsection{Core 1}
\label{subsec:core1}
\subsubsection{Physical evolution of Core 1}

Classically, core formation was assumed to be spherically symmetric in the absence of strong magnetic fields \citep{shu87}.  Later observations, especially with the Herschel telescope, found that molecular clouds are full of filamentary substructures and that dense cores are located along or at the intersections of filaments \citep{andre14}.  The view on the accretion process of dense cores has changed; for example, \citet{ren2021} found gas flow along a filament directly toward a pre-stellar core in OMC3 MMS-7.

\begin{figure*}
\includegraphics[scale=0.73]{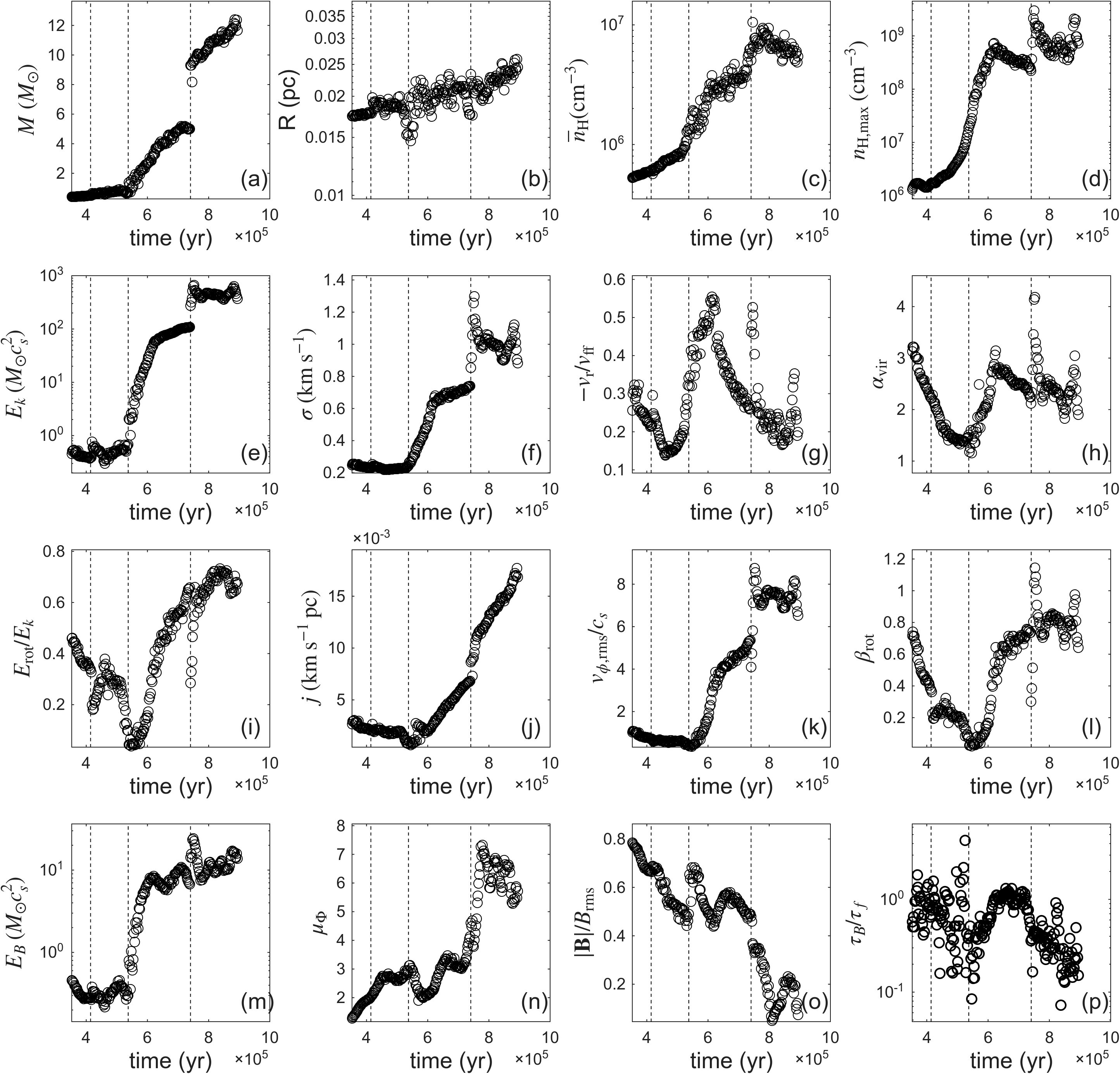}
\caption{Evolution history of physical properties of Core 1 for (a) total mass, $M$, (b) equivalent spherical radius of the core, $R$, (c) mean density, $\nbh$, (d) central density in the core, $n_{\rm H, max}$, (e) total kinetic energy, $E_K$, (f) 1D velocity dispersion $\sigma$, (g) ratio of inward velocity to the freefall velocity, $-\langle v_{\rm r} \rangle/\vff$, (h) virial parameter, $\alpha_{\rm vir}$, (i) ratio of rotational energy, $E_{\rm rot}$, to total kinetic energy, $E_K$, (j) specific angular momentum, $j$, (k) rotational velocity, $v_{\phi,\rm rms}$, in units of the sound speed, $c_s$, (l) $\beta_\rot$, a measure of the ratio of rotational energy to gravitational energy (eq. \ref{eq:beta}), (m) total magnetic energy, $E_B$, (n) normalized mass-to-flux ratio, $\mu_\Phi$, (o) ratio of magnetic field magnitude to the rms value, $|\mathbf{B}|/B_{\rm rms}$, and (p) the ratio of torque from gas inflow to the magnetic torque, $\tau_B/\tau_f$.  The three vertical dashed lines indicate significant mergers with massive gas clumps or cores during the evolution.  See the discussion in Section \ref{subsec:core1}. 
\label{fig:core1p}}
\end{figure*}

\begin{figure}
\includegraphics[scale=0.6]{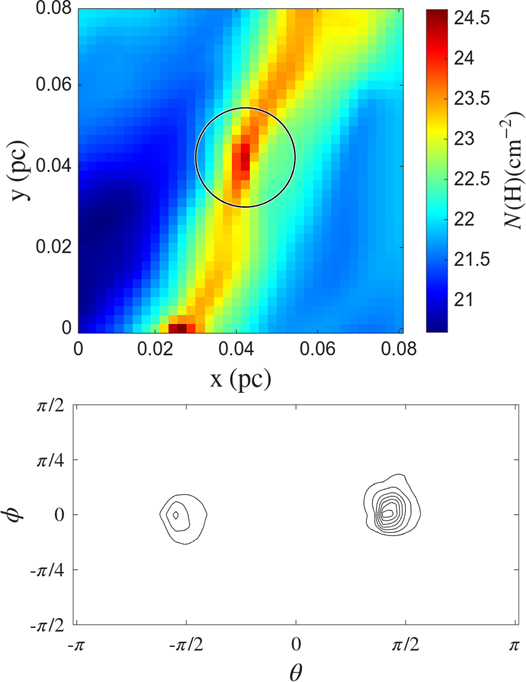}
\caption{Top: Column density map of Core 1 in a subfilament.
The circle indicates the spherical surface where the radial mass influx accretion rate is computed.  Bottom: A contour plot of the radial mass influx accretion rate projected from a spherical surface onto a Cartesian plane.  The contours are from 0.1 to 0.7 \msun\; \pc$^{-2}\; {\rm yr}^{-1}$, in increments of 0.1 \msun\; \pc$^{-2} {\rm yr}^{-1}$.  The two groups of contours indicate accretion along the subfilament in which the core is embedded.
\label{fig:massflux}}
\end{figure}

Figure \ref{fig:core1p} shows the evolution of the physical quantities of Core 1 beginning around $t=346$ kyr when it is about 0.4 \msun,  omitting the earlier times when the core was not distinct from the complex subfilaments in which it was embedded.  The first row of panels in the figure shows the evolution of the mass, radius and density. Core 1 grows by both accretion along one or more subfilaments in which the core is embedded (see Figure \ref{fig:massflux} and the discussion there) and by mergers with other cores.  In Figure \ref{fig:core1p}a, we see that there are several significant jumps in the relative mass due to episodic merging with subfilaments or another dense core.  The three dashed lines in the plots indicate three significant mergers--i.e., mergers that result in an increase in the current mass of the core of at least 20 per cent, even if the mass is small.  At the end of the simulation, 40\% of the final mass of Core 1 is from significant core mergers, with the final merger being a major merger.
The mass accretion rate of this core is indicated by the slope of $M(t)$ in Panel 15a.  On average (including significant mergers) the accretion rate is $2.1\times10^{-5}$ M${_\odot}$ yr$^{-1}$.  Between significant merger events, the accretion rate changes depending on the availability of gas nearby, with an average rate of $\dot M=1.26\times10^{-5}$ \msun yr\e.

The mean radius of the core, $R$, in Figure \ref{fig:core1p}b is obtained by assuming that the irregularly shaped core has a spherical volume. Knowledge of the mass and radius enables the determination of the mean density, which is shown in Figure \ref{fig:core1p}c.  Although there are significant fluctuations in the radius and density during the major mergers that occur in the evolution of Core 1, the mean density, $\nbh$, does not show systematic changes then  because the mean densities of the two merging cores are similar.  The central maximum density, $n_{\rm H,max}$, of Core 1 shows a large time variation near the last major merger in Figure \ref{fig:core1p}d.  The mean density and maximum density are much greater than one would expect for a singular isothermal sphere, $\nbh=3\nh(R)\simeq 10^6$ cm\eee\ and $n_{\rm H,max}=(R/\Delta x)^2\nh(R)\simeq 3\times 10^7$~cm\eee, respectively, indicating that in reality a star would  have already formed in this core.

Row 2 in Figures \ref{fig:core1p} shows quantities related to velocity. The total kinetic energy, $E_K$, includes both the rotational and infall energies.  It increases by more than 3 orders of magnitude after $t=600$ kyr and shows clear jumps at the time of major mergers.  The 1D velocity dispersion, which includes the thermal sound speed and is defined below Equation (\ref{eq:avir}), jumps significantly at the third merger.  In Figure \ref{fig:core1p}g, the ratio $\avg{v_r}/\vff$, shows the mass-averaged radial inward velocity of the core in units of the freefall velocity, $\vff=\sqrt{2GM/R}$, indicating that Core 1 is collapsing.

Panel \ref{fig:core1p}h shows the virial parameter, $\avir$, which is defined as the twice the ratio of the kinetic and gravitational energies, $E_K$ and $|E_G|$, of a uniform sphere, $\avir\equiv 5\sigma^2R/(GM)$ (Eq. \ref{eq:avir}).  A uniform sphere is in virial equilibrium if $\avir=1$ and is bound for $\avir<2$. Observe that the gravitational collapse of the core (Panel 15g) is occurring, despite the fact that the virial parameter, $\avir$, generally exceeds 2 after 600 kyr (Panel 15f). 
Defining $a$ as the ratio of the actual gravitational energy to that of a uniform sphere,
$a=| E_G |/(3GM^2/5R)$, implies
\beq
\avir=a\,\frac{2E_K}{| E_G |}=\frac{a}{1-P_0/\bar P},
\label{eq:avir2}
\eeq
where $P_0$ is the ambient gas pressure, $\bar P$ is the mean pressure in the gas sphere, and $a$ is a parameter of order unity that depends on the shape and density structure of the core \citep{bertoldi1992}. A core is bound if $\avir<2a$ and is in virial equilibrium if $\avir=a$.
The virial theorem implies that self-gravitating, unmagnetized cores are bound for $2E_k/|E_G|=(1-P_0/\bar P)^{-1}<1$, or $\bar P>2P_0$. 
\citep{bertoldi1992}.  As these authors noted, the ambient pressure, $P_0$, thus plays an important role in determining the value of $\avir$ and therefore the structure of the cloud. It follows that bound cores ($\bar P>2P_0$) can have $\avir>2$ provided $a>1$. For example, one can readily show that a singular isothermal sphere has $a=5/3$ and $P_0=\bar P/3$ so that it is bound with $\avir=5/2$.  
The energy associated with infall contributes $\avg{v_r}^2/3$\footnote{For the case in which the kinetic energy is in the form of turbulence and infall, it is $E_K=\frac 32 \sigma_{\rm turb}^2+\frac 12 \avg{v_r}^2=\frac 32(\sigma_{\rm turb}^2+\frac 13 \avg{v_r}^2)$. Note also that the observed line-of-sight velocity dispersion for pure infall is $\frac 13 \avg{v_r}^2$.} to the 1D velocity dispersion, $\sigma^2=\frac 13 v_\rms^2+\cs^2$, that enters the observed value of $\avir$; note that the large increase in $\avir$ at the time of the third merger in Panel 15h is correlated with the large increase in $\avg{v_r}/\vff$ in Panel 15g at that time.  If the energy associated with infall is removed since it does not contribute to supporting the core against gravity, then the effective value of $\avir$ is reduced, consistent with the fact that the core is undergoing gravitational collapse even when the observed $\avir>2.5$.  Finally, note that although there is a transient fluctuation in $\avir$ at the third merger, there is no jump. The gravitational energy scales with the kinetic energy after the merger.

The third row of panels in Figure \ref{fig:core1p} shows the evolution of the rotational state of Core 1.  Figure \ref{fig:core1p}i gives the ratio of the total rotational energy to the total kinetic energy.  Here the rotational energy is defined so as to exclude the small-scale azimuthal component of the turbulent energy,
\beq
E_{\rm rot}=\frac 12\int \bar\rho(r,z)v_\phi(r,z)^2 2\pi r dr dz, 
\eeq
where $\bar\rho(r,z)=(1/2\pi)\int \rho(r,\phi,z)d\phi$ and $v_\phi(r,z)=[1/2\pi\bar\rho(r,z)]\int \rho(r,\phi,z)v_\phi(r,\phi,z) d\phi$ is the mass-averaged azimuthal velocity relative to the angular momentum vector of the core.  For more than half the time, the rotational energy is more than half of the total kinetic energy, indicating that rotation plays a significant role in the dynamical state of the core.  The specific angular momentum, $j$, and the density-weighted, rms rotational velocity of the core in units of the sound speed, $v_{\phi,\rm rms}/\cs$ (see Figures \ref{fig:core1p}j and \ref{fig:core1p}k), both increase substantially after about 600 kyr.  The continuous increase in specific angular momentum is due to the highly non-radial accretion from the connected subfilaments (see animation A3).
In the discussion of the magnetic field below, we shall see that the angular momentum input is adequate to overcome magnetic braking.  The parameter $\beta_\rot$ (Figure \ref{fig:core1p}l), a measure of the ratio of rotational energy to gravitational energy \citep{goodman93}, is given by
\beq
\beta_\rot=\frac{E_{\rm rot}}{E_{\rm grav}}\bigg\vert_{\rm us}=\frac{\omega^2}{4\pi G\bar\rho}=\frac{\avir E_{\rm rot}}{2 a_I E_k},
\label{eq:beta}
\eeq
where the subscript ``us" denotes uniform sphere, $\omega^2=2E_{\rm rot}/I$, and $I=(2/5)a_I MR^2$. Here $a_I$ is the ratio of the actual moment of inertia of the core to that of a uniform sphere; it is about 1 at $t=400$~kyr, reaches a minimum of 0.19 after the last major merger, and returns to 0.57 at the end of the simulation. The final expression in Equation (\ref{eq:beta}) relates Panels 15h, i and $\ell$.  Averaging over the orientation of the rotation axis implies that the average observed value of $\beta_\rot$ is 2/3 of this \citep{chan50}.  A factor contributing to the decline in the infall velocity after 600 kyr (Panel g) is the increasing effect of rotation.  As in the case of the virial parameter, the third merger causes a fluctuation in $\beta_\rot$ but no jump: $\beta_\rot\propto E_\rot/E_G=(E_\rot/E_K)(E_K/E_G)$, and neither factor jumps at the merger.

The last row of panels of Figures \ref{fig:core1p} shows the parameters related to the magnetic field of Core 1.  The magnetic energy, $E_B$, in Figure \ref{fig:core1p}m has an unequivocal increase after the second significant merger.  This is also seen in other physical properties.  It is the result of large amounts of gas accreted from the attached subfilaments (see animation A3).  In Figure \ref{fig:core1p}n, the mass-to-flux ratio relative to the critical value, $\mu_{\Phi}=2\pi G^{1/2}M/\pi R^2 B_\rms$, fluctuates between 2 and 3.5 before the last major merger and then about doubles after that merger.  This is the result of magnetic reconnection just after the major merger with another magnetized core and the development of a strong magnetic toroidal component in the plane of the disklike core due to the fast rotation of Core 1 (see animation A3).  The ratio of the magnitude of the mean magnetic field to the rms value, $|\vecB|/B_\rms$, in Figure \ref{fig:core1p}o, shows that the mean magnetic field becomes an increasingly small fraction of the total field with time and is subdominant after the last major merger.  At late times, the mean field is less than 20\% of the rms field, indicating that Zeeman observations that did not resolve the core would significantly underestimate the field strength.

Finally, Figure \ref{fig:core1p}p gives the ratio of the torque due to magnetic braking to the torque from gas inflow to the core, ($\tau_B/\tau_f$), where \citep{kulsrud2005}
\beq
\mathbf{\tau}_B = \oint_S (\mathbf{r}\times\mathbf{B}) \frac{\mathbf{B}}{4 \pi} \cdot d\mathbf{S}
\eeq
and
\beq
\mathbf{\tau}_f = \oint_S \mathbf{v}\cdot d\mathbf{S} (\mathbf r \times \rho \mathbf{v}),
\eeq
where $\mathbf{B}$ and $\mathbf{v}$ are the magnetic and velocity vectors, respectively, and $\mathbf{r}$ is the position vector of a computational cell on the surface of the core,  $\mathbf{S}$, relative to the center of mass of the core.  Figure \ref{fig:core1p}p shows that the magnetic torque is generally less than the accretion torque, showing that magnetic braking is not strong enough to stop the rotation of Core 1.

In animation A3 (linked with Figure \ref{fig:helicalfield}), we show the formation and growth of Core 1 from the time we turn on gravity ($t=0$) to the end of the simulation ($t=894$ kyr), when the mass is 12.4 \msun.
In Figure \ref{fig:massflux}, we show the radial inward mass flux, $\rho v_r$, pointing to the center of Core 1 at time 688 kyr.  
For illustration, we calculated $\rho v_r$ on a radius of 0.02 pc spherical surface just outside Core 1 (top panel in Figure \ref{fig:massflux}).  The $\rho v_r$ through the spherical surface is projected onto a map shown in the bottom panel of Figure \ref{fig:massflux}.  The mass flux from the connected subfilaments is clearly indicated by the two groups of contours starting from 0.1 \msun\, pc$^{-2}$ yr$^{-1}$ with increments of 0.1 \msun\, pc$^{-2}$ yr$^{-1}$.  The rest of the spherical surface has $\rho v_r < 0.1$ \msun\, pc$^{-2}$ yr$^{-1}$.  As noted above, the average accretion rate of Core 1 is about $2.1\times10^{-5}$ \msun yr$^{-1}$.  The total $\rho v_r$ inside the areas enclosed by the contours at 0.1 \msun pc$^{-2}$ yr$^{-1}$ is 1.4 times the mass flux from the rest of the spherical surface, whereas the area of the two groups of contours is only 7.4\% of the total area of the spherical surface.  Therefore, the mass accreted onto the core is highly concentrated from the connected subfilament and so is likely to penetrate more deeply into the core.

Our results are consistent with those of \citet{smith2011}, who used numerical simulation to examine the accretion of gas onto protostars embedded in filaments.  They found that protostars accrete first from the core and then from outside the core, primarily through the filament in which the core is embedded.  It should be noted that our results pertain to an earlier time, when the core is forming.

\subsubsection{Shocks around subfilaments and Core 1}

Supersonic turbulence in molecular clouds cascades from the largest scales down to the sonic scale.  In Section \ref{subsec:filament_form}, we discussed the formation of small filaments due to shock compression by supersonic turbulence.  In Section \ref{sec:filament_stab}, we discussed how the collision of subfilaments inside molecular clouds could result in core formation.  When there is supersonic motion, shocks are expected with large increases in density.  Here we use Core 1 as an example to discuss the effect of supersonic turbulent flows on the assembly of a core.

\begin{figure*}
\includegraphics[scale=0.94]{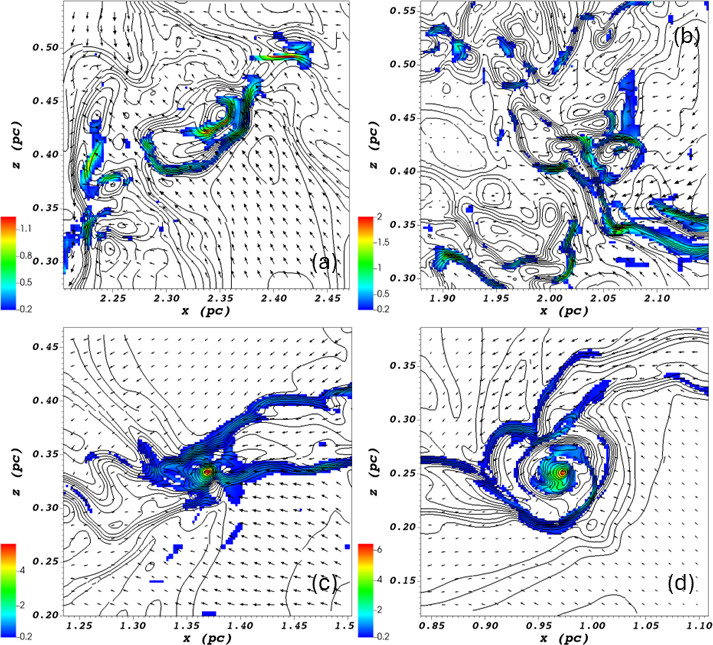}
\caption{Shocks in a supersonic region with a size in each dimension of 0.27 pc around Core 1.
The panels are comoving with the gas, primarily to the left.  The color map is $-\Delta x\nabla \cdot ({\mathbf v}/\cs)$ plotted from 0.2 in the plane of the slides at times (a) $t = 0$, (b) $t = 194$ kyr, (c) $t = 660$ kyr, and (d) $t = 894$ kyr.  Thirty black contours show the volume number density on a logarithmic scale separated by 0.2 dex from $\nh=10^2$ to $10^8$ cm$^{-3}$.  The maximum densities at these times are $3.6\times10^5$, $2.1\times10^6$, $4.3\times10^8$, and $7.3\times10^8$ cm$^{-3}$, respectively.  Black arrows show the direction of gas flow in the plane of the slides. 
\label{fig:core1_shocks}}
\end{figure*}

The best indicators of a shock are jumps in density and velocity.  Here we shall use the velocity jump measured by the negative of the normalized 3D divergence of the Mach number, $-\Delta x\nabla\cdot({\mathbf v}/\cs)$, where $\Delta x = 2.2 \times 10^{-3}$ pc is the size of the computational cell at the highest level of refinement.  In Figure \ref{fig:core1_shocks} we show a sequence of two-dimensional slides of the divergence of the normalized velocity in color together with density contours in a region of 0.27 pc where the subfilaments and a core will form in time. It is the same region shown in Figure \ref{fig:helicalfield}, where a helical field appears between 170 - 220 kyr.
Since the gas is compressed where shocks occur, we show only the negative values of $\nabla \cdot ({\mathbf v}/\cs)$.  In fact, to visualize the correlation between density compressions and shocks, the spacing of contour lines is more useful than the density values.  The velocity vectors are superimposed on the map to show the directions of the turbulent gas flow in the slides.

Figure \ref{fig:core1_shocks}a shows the time when gravity has just turned on.  The dense structures at this time are solely the result of supersonic compression in a magnetized, turbulent environment.  The shock locations generally match the large density gradients well.  Note that $-\Delta x\nabla \cdot ({\mathbf v}/\cs)$ is plotted with a minimum value of 0.2.  At a value of 0.5, the magnitude of the velocity jump in $2\Delta x$ is equal to the sound speed. Anything below 0.2 is in white color.  Also note that the color bar scales vary with time as the shocks become stronger.  Shocks appear at many locations, as expected in a supersonically turbulent system.  In this small region, much of the gas is being assembled together by supersonic gas flows.  In Figure \ref{fig:core1_shocks}b, 194 kyr after gravity is switched on, shock structures still appear randomly, and Core 1 has not yet clearly formed.  At $t= 660$ kyr (panel c) there is a well-defined subfilament and the Core 1 confined by the compressive flows.  Gravity dominates the turbulence in the subfilament and the embedded core at this time (see also Figure \ref{fig:core1_energy}).  The subfilament is mainly held together by self-gravity with the help from shock pressure due to supersonic flows hitting the surface of the dense gas subfilament.  In this case, the ambient pressure that contributes to the confinement of the cloud (Eq. \ref{eq:avir2}) is anisotropic. Panel (d), at the end of the simulation (894 kyr), shows that the main core has merged with another core, resulting in a rapidly rotating disklike structure with spiral-like shock structures in the disk.  At both 660 kyr and 894 kyr, the very large value of $-\Delta x\nabla \cdot ({\mathbf v}/\cs)$ in the core is due to gravitational collapse.  
There is observational evidence of supersonic shocks near the surface of subfilaments and cores \citep[e.g.][]{dickens2001,soma2015,scibelli2024,hsu2025}.

\subsubsection{Energy balance around Core 1}

\begin{figure*}
\includegraphics[scale=0.8]{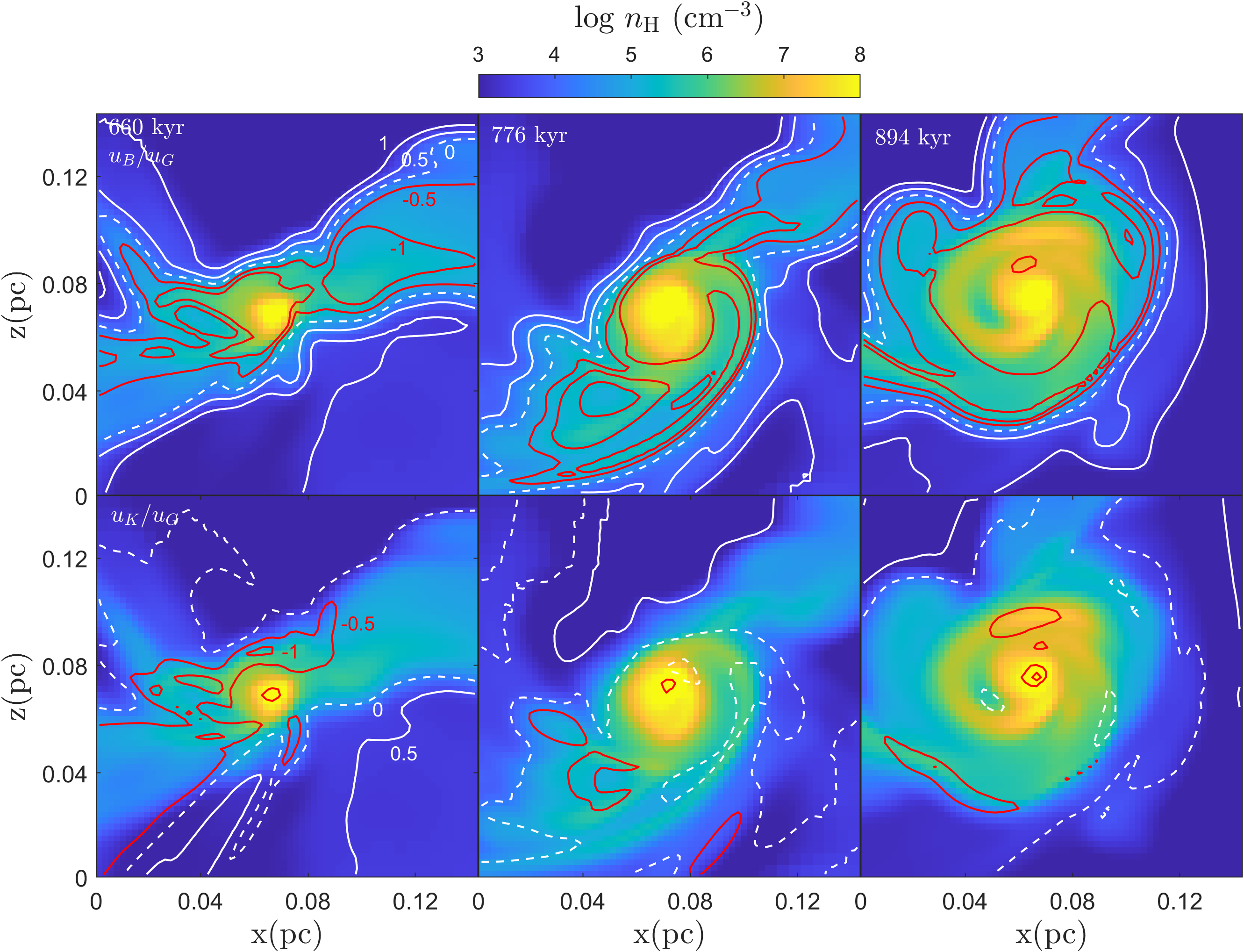}
\caption{Top row: slices, through the center of Core 1, of the ratios of magnetic to gravitational energy density on a logarithmic scale, $\log(u_B/u_G)$, 
at times 660, 776, and 894 kyr from left to right.  
The white dashed contour corresponds to $u_B/u_G=1$.  The red contours correspond to $\log(u_B/u_G)= -0.5$ and -1 (gravitationally dominated), and the white contours to $\log(u_B/u_G)=0.5$ and 1 (magnetically dominated).  The log-scale values of the ratios for the contours are shown in the top- and bottom-left panels as example. The color image is a slice of the volume density, $\nh$.  The size of the region is $0.15\times0.15$ pc.  The core 1 is located at the center of each figure panel.  Bottom row: same as top row for the ratio of kinetic energy to gravitational energy density, log$(u_K/u_G)$.
\label{fig:core1_energy}}
\end{figure*}
In Figure \ref{fig:core1_energy}, we show the ratios of magnetic and kinetic energy density, $u_B$ and $u_K$, respectively, to the gravitational energy density, $u_G$, at each cell at different times from 660 kyr to 894 kyr in a smaller region of 0.15 pc to observe the gravitationally dominated zones around Core 1.  The kinetic energy density is
$u_K = 0.5~\rho_{\rm cell}~(v^2+3c_s^2)$, where $\rho_{\rm cell}$ is the density of the computational cell and $v$ is the velocity of the cell in the rest frame of the region shown in the figure.  The gravitational energy density is computed as $\rho_i \varphi_i$ for each cell $i$ in the core, where $\varphi_i$ is the gravitational potential in cell $i$.  The gravitational potential $\varphi_i$ is derived from the density via the Poisson equation over the full simulation volume.  The dashed contour in each figure panel shows the location where the two energies in the ratio are equal.  Inside that contour, the gravitational energy exceeds the kinetic energy, in seeming contradiction to Figure \ref{fig:core1p}, which shows $\avir\simeq 2 E_K/|E_G|\ga 2$. However, the gravitational energy in Figure \ref{fig:core1_energy} is due to all the mass in the simulation volume, whereas that in the virial parameter is due only to the mass in the core.  The core is rotating about an axis approximately in the $y$ direction, which is normal to the plane of the figure.  The shape of the gravitation-dominated region enclosed inside this contour is highly irregular as a result of turbulence, which is highly supersonic (Fig. \ref{fig:core1p}).  In the case of $u_B/u_G$, the surface of equal energies is usually located at $\nh\sim 10^{4-5}$ cm$^{-3}$, outside the surface of $3\times10^5$ cm$^{-3}$, where we define a core for analysis; gravity therefore dominates magnetic fields in this core \citep{crut12}.  The size of the region inside the equal-energy contours around the dense core increases in time (from 776 to 894 kyr) as more gas mass accretes onto the dense core indicating the increasing influence of gravity over other forces with time.  
From the evolution history of Core 1, we see that core mergers can effectively accelerate the growth of cores in dense regions.  Subfilament collisions and core mergers naturally lead to the formation of hub-filament systems at core scales similar to the hub-filament systems on clump/cloud scales that can lead to the formation of massive stars \citep[e.g.][]{myers2009,trevino2019,kumar2020,zhou2022}.

\begin{figure}
\begin{interactive}{animation}{A4.mp4}
\includegraphics[scale=0.2]{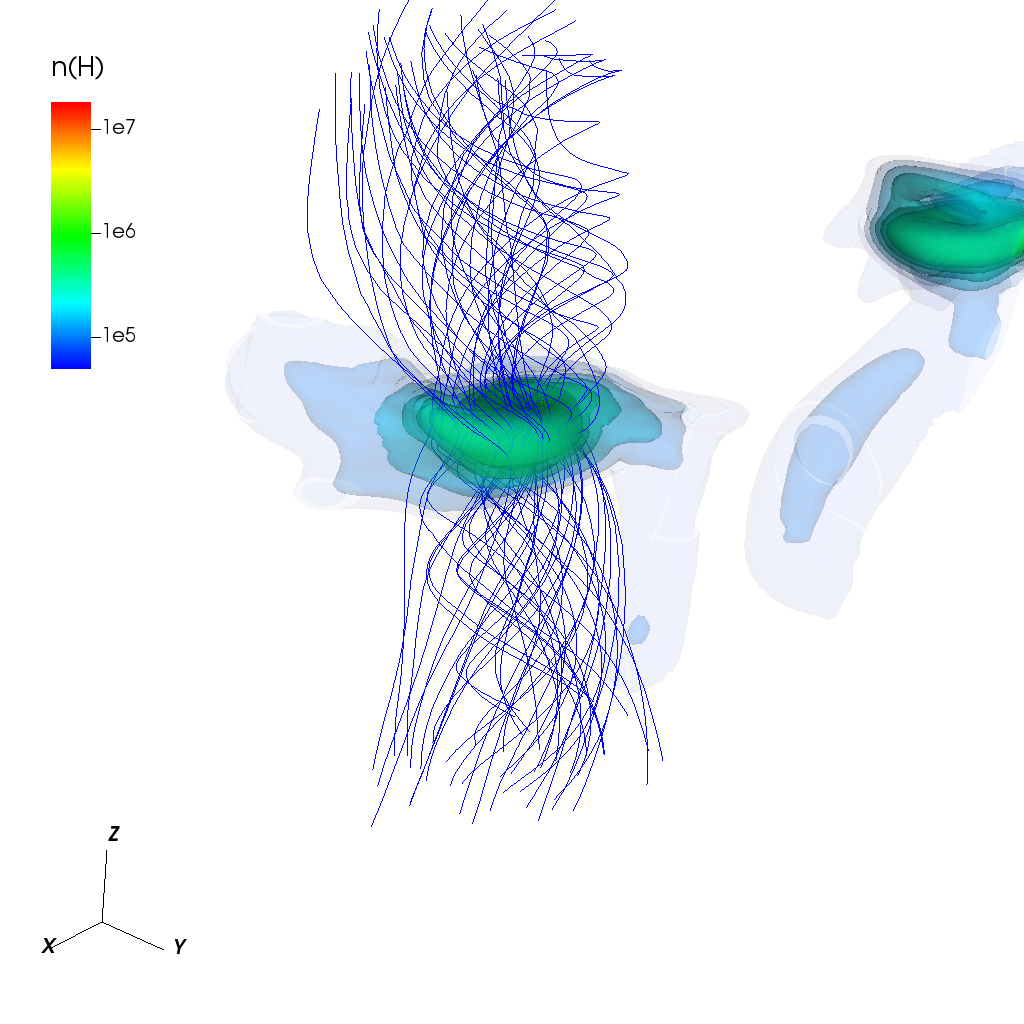}
\end{interactive}
\caption{Magnetic field structure of Core 2 at the end of the simulation.  The animation shows the evolution of Core 2 with magnetic field in the entire simulation.  \it{The animation associated with this figure is available in the online version of the published article on ApJ.}
\label{A4}}
\end{figure}

\begin{figure*}
\includegraphics[scale=0.73]{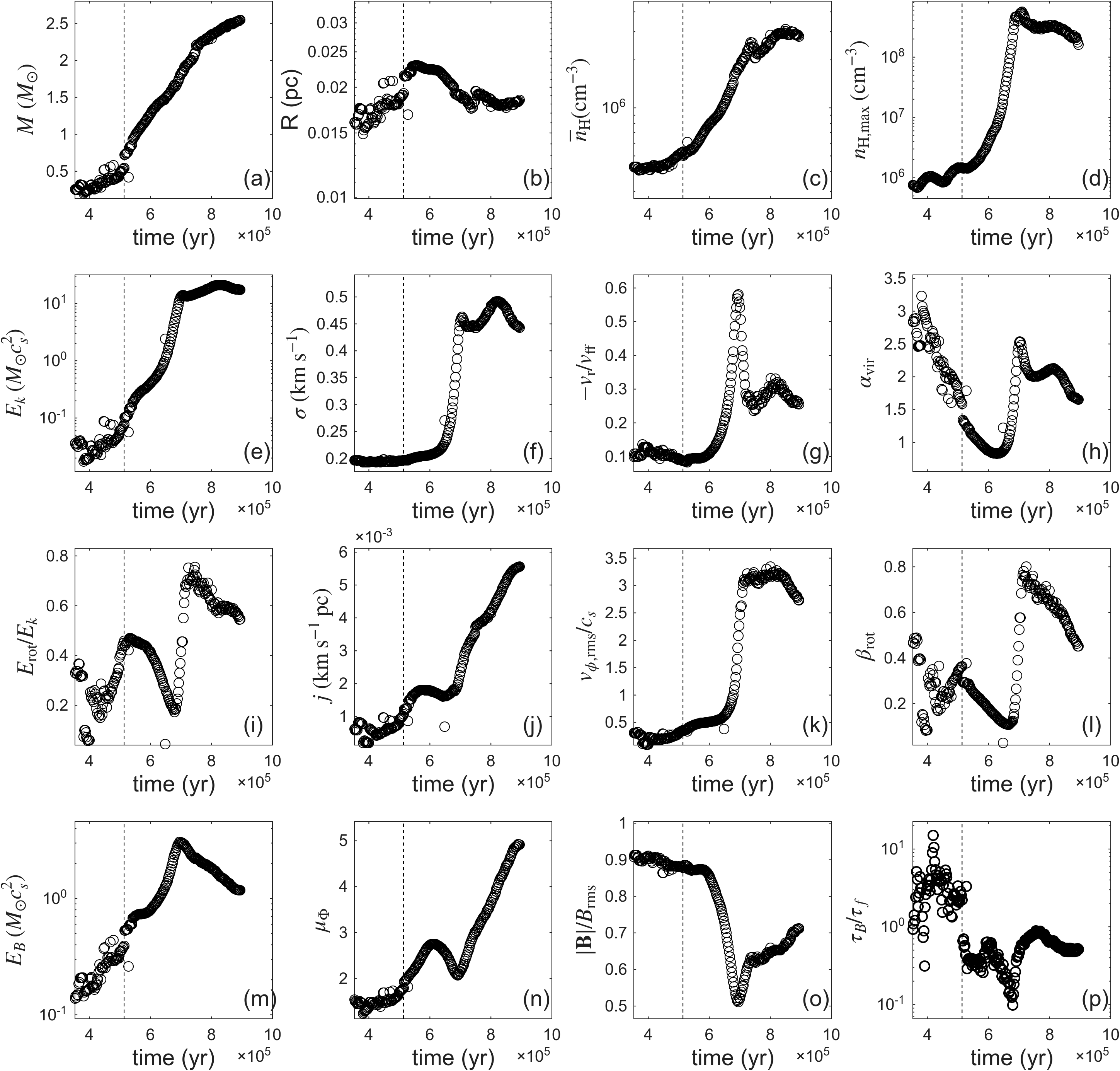}
\caption{Same as Figure \ref{fig:core1p} for Core 2. The vertical dashed line indicates a significant merger with gas clump less than 50\% of the mass of Core 2.  See the discussion in Section \ref{subsec:core2}.
\label{fig:core2p}}
\end{figure*}

\subsection{Core 2}
\label{subsec:core2}

The animation A4 (Figure \ref{A4}) shows the evolution of another core, Core 2, which evolves in a less crowded region in the simulation and ends with a smaller mass than Core 1, 2.6 \msun.  Figure \ref{fig:core2p} shows the evolution history of the physical properties of this core.  Before 200 kyr, we see several subfilaments in the region.  They interact and merge into a rotating, flattened gas structure of about 0.04 pc in size, with an FWHM $W=0.025$~pc.  A clear dense core (gas with $\nh>3\times 10^5$~cm\eee) does not appear until around 400 kyr.  The core is still attached to a thin subfilament of width about 1/3 of the core diameter.  After 600 kyr, the flattened gas structure contracts onto the core, smoothly increasing the mass (Figure \ref{fig:core2p}a).  The contraction rapidly accelerates (Figure \ref{fig:core2p}g) until it is curtailed by the increase in rotational velocity (Figure \ref{fig:core2p}k), which causes an increase in $\beta_{\rm rot}$ (Figure \ref{fig:core2p}l).  The magnetic field first develops a helical form near the center, where the rotation rate is a maximum (Figure \ref{A4}).  The rotating dense core evolves into a toroidal shape and has a small dip in the density isosurface at the center.

Compared to Core 1, the evolution of Core 2 is smoother after initially mingling with connected subfilaments and the merger with a gas structure at around 513 kyr.  The mass fluctuation in Figure \ref{fig:core2p}a before 513 kyr is the result of the uncertainty in determining the center of the core when it is still connected to many gas structures.  The average accretion rate of the core is $4.5 \times 10^{-6}$ M${_\odot}$ yr$^{-1}$, with a dispersion of $8.6 \times 10^{-7}$ M${_\odot}$ yr$^{-1}$.  The average rate is about 1/5 that of Core 1.  The difference comes mainly from the major merger of Core 1 with a core of almost equal mass at 741 kyr.  In fact, there is a dense core near Core 2, but it never comes close enough to merge during the entire simulation.  The value of $\alpha_{\rm vir}$ is generally below 2, indicating  that the core is gravitationally bound. 
As for Core 1, this is consistent with the fact that the core is undergoing gravitational collapse (Fig. \ref{fig:core2p}h). The value of $\tau_B/\tau_f$ remains below unity most of the time, indicating that the magnetic field cannot stop the rotation.  The value of $j$ of Core 1 is more than three times that of Core 2 as the result of several major mergers experienced by Core 1.  The specific angular momenta of cores in the filamentary cloud L1495/B213 are found to be in the range $0.2-7\times10^{-3}$ km s$^{-1}$ pc by \citet{punanova2018}.  For Core 2, we can see in Figure \ref{fig:core2p}d that $j$ increases steadily with time over the range $0.5 - 5.5\times10^{-3}$ km s$^{-1}$ pc, similar to the cores in L1495/B213.  However, for Core 1, $j$ is about $7\times10^{-3}$ km s$^{-1}$ pc right before the last major merger and grows to more than twice that by the end of the simulation.  This simulation shows that gas inflow through attached gas subfilaments and mergers with gas clumps or cores are very effective mechanisms for maintaining a rotating disk-like gas core.  From the evolution animations of Core 1 and 2, we can see the disklike geometry can be maintained for long periods of time: That for Core 1 lasts for at least 200 kyr, and that for Core 2 lasts for about 100 kyr to the end of the simulation.

\section{Conclusions}
\label{sec:conclusion}

Since the Herschel observations, we have extensive observational information about the filamentary structures both in and of molecular clouds \citep{hacar23,pineda23}.  However, limited resolution, the lack of full 3D information, and the long lifetimes of clouds and cores all interfere with our ability to determine how filaments form and evolve in molecular clouds. Observations show that molecular clouds have a hierarchical structure, with cores inside one or more levels of clumps within GMCs.  In a nearby GMC such as Taurus, we see that the cores are embedded in fibers, which are embedded in filaments that make up the overall filamentary molecular cloud.
In this work, we have analyzed results from the simulation of a filamentary molecular cloud (Paper II) to visualize and understand how subfilaments, the computational analogs of fibers, form, how they evolve, and how dense cores form and evolve along subfilaments.  We find that:

\begin{itemize}
\item[1] {\it Formation of the main filament.}
As reported in Paper II, many subparsec-long filaments arise in the simulation before gravity is activated.  As shown here, the large majority of these filaments are the result of compression by shocks moving at an angle with respect to the magnetic field.  After gravity is turned on, the sub-parsec filaments are assembled by large-scale turbulent flows with a significant assist from gravity to form a long, massive filament extending across the simulation volume and containing about 15 percent of the mass in the simulation.  This main filament has a FWHM $\simeq 0.075$ pc within the range of filament FWHMs observed by \citet{arzoumanian19}.

\item[2.] {\it Formation of subfilaments.}
After gravity is turned on, the small, subparsec filaments grow in both line mass and central density and become subfilaments in the main filament.  The growth in line mass is due both to mergers with other subfilaments and to accretion of diffuse gas; the growth in central density is due to compression in the main filament and to self-gravity in the subfilament.  The growth of filaments by mergers of subfilaments in our simulation confirms the proposal by \citet{hacar13}.

\item[3.] {\it Physical properties of subfilaments identified by $\get$ from the projected column density map.}
The median line mass and FWHM of subfilaments identified by $\get$ are $\sim 11 - 37$ \msun\ pc\e\ and $\sim 0.03 - 0.04$ pc, respectively, near the end of the simulation.  The median central column density is
$\Nhc \sim 1 \times 10^{22}$ cm$^{-3}$.  The maximum $\Nhc$ is over $10^{23}$ cm$^{-2}$.  The beam size used to smooth the simulation data and the characteristic width, $\wch$, used in the $\get$ analysis affect  subfilament identification lead to variations in the median values of the physical properties of subfilaments, just as in the case of actual observations \citep[e.g.][]{kirk2015}.  The physical properties of the subfilaments are similar to those of the ``fibers" observed in Orion by \citet{hacar18}, while the properties of the main filament are similar to those of the filaments observed by \citet{arzoumanian19}.

\item[4.] {\it Core formation.} 
We find that the dominant process of core formation begins with mergers of subfilaments. Subcritical subfilaments (i.e., subfilaments that are stable against radial gravitational collapse) merge and become supercritical subfilaments  (what \citealp{hacar13} called ``fertile" filaments).  In the simulation, the two subfilaments we examined in detail form cores at locations where the line mass exceeds the critical value—a sign of instability—with one producing a single core and the other a chain of cores, as the subfilaments continuously merge with nearby subfilaments.  We tracked the evolution of a total of nine randomly chosen cores with masses larger than 1 \msun\ at the end of
the simulation; all formed as the result of collisions and mergers of subfilaments. Colliding subfilaments and the core formed at their intersection form a hub-filament system similar to the hub-filament systems on clump/cloud scales associated with the formation of massive stars, but on a smaller scale \citep[e.g.][]{zhou2022}.

\item[5.] {\it Core evolution.}
For the nine cores we studied, with two shown in detail, the angular momentum of the cores increased via highly asymmetrical accretion, supplied by both misaligned, attached subfilaments and the surrounding gas reservoir. The non-radial infall trajectories exerted a net external torque on the cores, which was able to overcome magnetic braking.  Whereas the classical theory of low-mass star formation assumes that pre-stellar and protostellar cores grow by accretion of diffuse gas \citep{shu87}, we found that mergers with other cores can also be significant. It is possible that such mergers play a role in the formation of binary stars; higher resolution simulations would be necessary to determine that.  In an examination of nine cores larger than 1 \msun, four out of the nine experienced at least one major merger, defined as an increase in the mass of the core of at least 50 per cent; such cores had larger final masses than cores that did not undergo major mergers. The median accretion rate of the four cores with a major merger is
$2.3\times10^{-5}$ \msun\ yr\e; that of the five cores that did not experience a major merger is $1.0\times10^{-5}$ \msun\ yr\e.
The typical accretion time scale for the cores exceeds $5\times 10^5$ yr.

\item[6.] {\it From Turbulent Confinement to Gravitational Dominance.}
During the evolution of dense substructures inside the main filament, shocks are frequently observed around dense subfilaments and cores \citep[e.g.][]{hsu2025}.  Supersonic turbulent flows provide crucial pressure confinement to dense gas structures at early times until gravity dominates and leads to gravitational collapse of cores.

\end{itemize}

\begin{acknowledgments}
We thank Alexander Mensh'chikov for his help in using the \get\ code for subfilament identification in this paper.  We are grateful to the referee for the careful review and helpful suggestions, which greatly improved the quality of this paper.  PSL acknowledges the support by the National Key R\&D Program of China (No. 2022YFA1603100), National Natural Science Foundation of China (NSFC) (No. 1241101426), the Natural Science Foundation of Shanghai (No. 23ZR1482100), the Strategic Priority Research Program of the Chinese Academy of Sciences (CAS) Grant No. XDB0800303, and the PIFI program of Chinese Academy of Sciences through grant No. 2025PG0009.  This research used resources of the National Energy Research Scientific Computing Center (NERSC), a Department of Energy Office of Science User Facility using NERSC award NP-ERCAP0029603 and NP-ERCAP0031919.
\end{acknowledgments}

%

\vspace{5mm}


\software{ORION2 \citep{li12a,li21},  
          }



\appendix

\section{The method of tracing the evolution of cores}

In this section, we describe how we monitor the evolution  of cores as they grow due to accretion of gas from the subfilaments with which they are connected and from mergers with nearby cores.  Tracking the movement of a core in a simulation is not trivial \citep[e.g.][]{offner2022}, especially using a grid-based code like \textsc{Orion2}.  In order to follow the evolution of a core, we first identify it at the end of the simulation, when cores are well defined with very dense cells near the center.  We use CLUMPFIND to identify its extent in every data dump and trace its evolution backward in time until we cannot clearly identify it.  The coordinates of the densest cell in each core are recorded.  We then load the density data from the previous time step and locate the $11\times11\times11$ region with the densest cell at the center at the densest cell.  In this region, we find the highest density cell and use CLUMPFIND to identify the extent of the core.  Unless the core is moving extremely fast, the densest cell will still be in the square box at the previous time step, and the box is re-centered.  The size of this tracking region is sufficient in our nine-core study.  We continue this process to track the movement of the core center backward in time until the density of the densest cell is comparable to that of the nearby gas structures.  When two cores merge, we track the one with higher mass backward in time.  As seen in the animations A1 and A2, the cores are connected with subfilaments most of the time.  To monitor the formation of a core inside a subfilament, we include only gas with a density above the threshold for a core, $\nthr = 3\times10^5$ cm$^{-3}$; the gas becomes part of an actual core only when its density also exceeds that of the ambient medium.


\bibliography{references}{}
\bibliographystyle{aasjournal}



\end{document}